\documentclass[bibyear]{aa}
\usepackage[varg]{txfonts}
\usepackage{graphicx}
\usepackage{longtable}
\usepackage{lscape}
\begin{document}

\title{Gaia-ESO Survey: Gas dynamics in the Carina Nebula through
optical emission lines.
\thanks{
Based on observations collected with the FLAMES spectrograph at VLT/UT2
telescope (Paranal Observatory, ESO, Chile), for the Gaia-ESO Large
Public Survey (program 188.B-3002).
}
}

\date{Received date / Accepted date}

\author{F. Damiani\inst{1},
R. Bonito\inst{1,2}, L. Magrini\inst{3}, L. Prisinzano\inst{1},
M. Mapelli\inst{4},
G. Micela\inst{1},
V. Kalari\inst{5,6},
J. Ma\'iz Apell\'aniz\inst{7},
G. Gilmore\inst{8},
S. Randich\inst{3},
E. Alfaro\inst{9},
E. Flaccomio\inst{1},
S. Koposov\inst{8,10},
A. Klutsch\inst{11},
A.~C. Lanzafame\inst{12},
E. Pancino\inst{3,13,14},
G.~G. Sacco\inst{3},
A. Bayo\inst{15},
G. Carraro\inst{16},
A.~R. Casey\inst{8},
M.~T. Costado\inst{9},
E. Franciosini\inst{3},
A. Hourihane\inst{8},
C. Lardo\inst{17},
J. Lewis\inst{8},
L. Monaco\inst{18},
L. Morbidelli\inst{3},
C. Worley\inst{8},
S. Zaggia\inst{19},
T. Zwitter\inst{20},
\and
R. Dorda\inst{21}
}
\institute{INAF - Osservatorio Astronomico di Palermo G.S.Vaiana,
Piazza del Parlamento 1, I-90134 Palermo,
Italy \\
\email{damiani@astropa.inaf.it}
\and
Dipartimento di Fisica e Chimica, Universit\`a di Palermo,
Piazza del Parlamento 1, 90134, Palermo, Italy
\and
INAF - Osservatorio Astrofisico di Arcetri, Largo E. Fermi
5, 50125, Firenze, Italy
\and
INAF - Osservatorio Astronomico di Padova, Vicolo dell'Osservatorio 5,
I-35122, Padova, Italy
\and
Armagh Observatory, College Hill, Armagh BT61 9DG, UK
\and
School of
Mathematics \& Physics, Queen's University Belfast, Belfast BT61 7NN, UK
\and
Centro de Astrobiolog\'ia (CSIC-INTA), ESAC campus, Camino bajo del
castillo s/n, 28 692 Villanueva de la Ca\~nada, Madrid, Spain
\and
Institute of Astronomy, University of Cambridge, Madingley Road,
Cambridge CB3 0HA, UK
\and
Instituto de Astrof\'isica de Andaluc\'ia-CSIC, Apdo. 3004, 18080,
Granada, Spain
\and
Moscow MV Lomonosov State University, Sternberg Astronomical Institute,
Moscow 119992, Russia
\and
INAF - Osservatorio Astrofisico di Catania, via S. Sofia 78, 95123,
Catania, Italy
\and
Dipartimento di Fisica e Astronomia, Sezione Astrofisica, Universit\`{a}
di Catania, via S. Sofia 78, 95123, Catania, Italy
\and
INAF - Osservatorio Astronomico di Bologna, via Ranzani 1, 40127,
Bologna, Italy
\and
ASI Science Data Center, Via del Politecnico SNC, 00133 Roma, Italy
\and
Instituto de F\'isica y Astronomi\'ia, Universidad de Valparai\'iso, Chile
\and
European Southern Observatory, Alonso de Cordova 3107 Vitacura, Santiago
de Chile, Chile
\and
Astrophysics Research Institute, Liverpool John Moores University, 146
Brownlow Hill, Liverpool L3 5RF, UK
\and
Departamento de Ciencias Fisicas, Universidad Andres Bello, Republica
220, Santiago, Chile
\and
INAF - Padova Observatory, Vicolo dell'Osservatorio 5, 35122 Padova,
Italy
\and
Faculty of Mathematics and Physics, University of Ljubljana, Jadranska
19, 1000, Ljubljana, Slovenia
\and
Departamento de F\'{\i}sica, Ingenier\'{\i}a de Sistemas y Teor\'{\i}a
de la Se\~nal, Universidad de Alicante, Apdo. 99, 03080 Alicante, Spain
}

\abstract
{}
{We present observations from the Gaia-ESO Survey in the lines of
H$\alpha$, [N II],
[S II] and He I of nebular emission in the central part of the Carina Nebula.
}
{We investigate the properties of the two already known kinematic
components (approaching and receding, respectively), which account for
the bulk of emission. Moreover, we investigate the features of the much
less known low-intensity high-velocity (absolute RV $>$50 km/s) gas emission.
}
{We show that gas giving rise
to H$\alpha$ and He I emission is dynamically well correlated, but not identical,
to gas seen through forbidden-line emission. Gas temperatures are derived
from line-width ratios, and densities from [S II] doublet ratios. The spatial
variation of N ionization is also studied, and found to differ between the
approaching and receding components.
The main result is that the bulk of the emission lines
{in the central part of Carina arises from several distinct shell-like
expanding regions}, the most evident found around $\eta$ Car, the Trumpler~14
core, and the star WR25. Such ``shells" are non-spherical, and show distortions
probably caused by collisions with other shells or colder, higher-density gas.
Part of them is also obscured by foreground dust lanes, while only very
little dust is found in their interior.
Preferential directions, parallel to the dark dust lanes, are found in
the shell geometries and physical properties, probably related to strong
density gradients in the studied region.
We also find evidence that the ionizing flux emerging from $\eta$ Car
and the surrounding Homunculus nebula varies with polar angle.
{The high-velocity components in the wings of H$\alpha$ are found to arise
from expanding dust reflecting the $\eta$ Car spectrum.}
}
{}

\keywords{ISM: individual objects: (Carina Nebula)
-- ISM: general -- HII regions
}

\titlerunning{Gas dynamics in Carina Nebula}
\authorrunning{Damiani et al.}

\maketitle

\section{Introduction}
\label{intro}

The Carina Nebula is one of the largest known star-forming complexes in
the Galaxy, and has been extensively studied thanks to its relatively
small distance (2.25$\pm$0.18~kpc,
Davidson and Humphreys 1997; 2.35$\pm$0.5~kpc, Smith 2006),
and moderate foreground reddening ($E(B-V)=0.36$, Hur et al.\
2012). A review of its properties is given by Smith and Brooks (2008).
Several young clusters are embedded in the Nebula,
most notably Trumpler~14 and 16, and Collinder~228, which all together form
the Car~OB1 association. This is one of the largest OB associations
in the Galaxy, with more than 60 stars earlier than B0, and several WR
stars. The most studied member of Trumpler~16 is the luminous blue variable
(LBV) $\eta$~Car.
About one-third of all known O3 stars in the Galaxy, and the first
discovered O2 star, are found in Trumpler~14 and~16.

The age of the Carina star-forming region (SFR) and its component clusters
has been estimated in
the range from 1 to several Myr. Wolk et al.\ (2011) remarks however that
no simple age sequence is able to explain the respective properties of
clusters Trumpler~14, 15, and~16, so that other parameters come into play to
define the properties of these regions. There are also hints of a past
supernova explosion in the region (see Smith and Brooks 2008,
Townsley et al.\ 2011).
The region is still actively forming stars in its outer parts (Smith et
al.\ 2000;
Smith, Bally and Brooks 2004; Smith, Stassun and Bally 2005; Povich et
al.\ 2011),
and shows many dense dusty patches (``pillars"), seen
against the very bright emission from the H$_{\rm II}$ region.
Extinction within the molecular cloud, where Carina star clusters are still
partially embedded, spans a wide range (up to $A_V \sim 15$ among X-ray detected
2MASS sources, and perhaps more, Albacete-Colombo et al.\ 2008),
hiding most background stars from
optical observations in the central parts of the clusters.

The stellar population in the Carina SFR was recently observed as part
of the Gaia-ESO spectroscopic Survey, which will cover more than $10^5$
stars belonging to all components of the Milky Way, including some very
young clusters (Gilmore et al.\ 2012, Randich et al.\ 2013).
More than 1000 low-mass stars, and several hundreds OB stars, have
been observed towards Carina, using the ESO VLT/FLAMES multi-fibre
spectrograph (Pasquini et al.\ 2002).
This allows simultaneous observations of $\sim 130$ targets (stars and sky
positions) at intermediate resolution ($R \sim 15000-20000$, depending on
setup) with the Giraffe spectrograph, and 7-8 positions at high
resolution with the UVES spectrograph. Several fibres per Observing
Block (OB; each OB comprises all spectra recorded simultaneously) were
aimed at ``empty" sky positions, to estimate the sky contribution
expected in stellar spectra. Therefore, the Survey data contain a rich
dataset of pure sky spectra, at positions scattered
across the Nebula, containing an amount of information on the ionized
component of the diffuse medium towards that line of sight, which we
study in this work. The spectra of low-mass stars in the field will be
studied in a later work (Damiani et al., in preparation).

The diffuse medium in Carina is already known to be complex, and contains
a mixture of components across a huge range of temperatures. For
example, cold gas was imaged in CO lines by Yonekura et al.\ (2005);
intermediate-temperature gas emitting in the optical lines was studied by
Deharveng and Maucherat (1975), Walborn and Hesser (1975),
Meaburn, Lopez and Keir (1984);
the whole Nebula was found to be permeated by
million-degree hot gas by Seward et al.\ (1979) and Townsley et al.\ (2011).
Smith and Brooks (2007) made a comparative review of large-scale
observations of its diffuse medium.

In this paper, we focus on the analysis of pure sky spectra in the
central part of the Carina nebula, obtained by the Gaia-ESO survey. In
particular, we focus on the kinematic properties of H$\alpha$, [NII], [SII]
and He~I nebular emission lines. In Section~\ref{data} we describe the
observations, while in Section~\ref{results} we present our results.
In Section~\ref{discuss} we
discuss the main implications of our work for the dynamics of gas in the
Carina nebula, while Section~\ref{summary} summarizes our conclusions.

\section{Observational data}
\label{data}

\begin{figure*}
\includegraphics[bb=0 0 756 761,width=18cm]{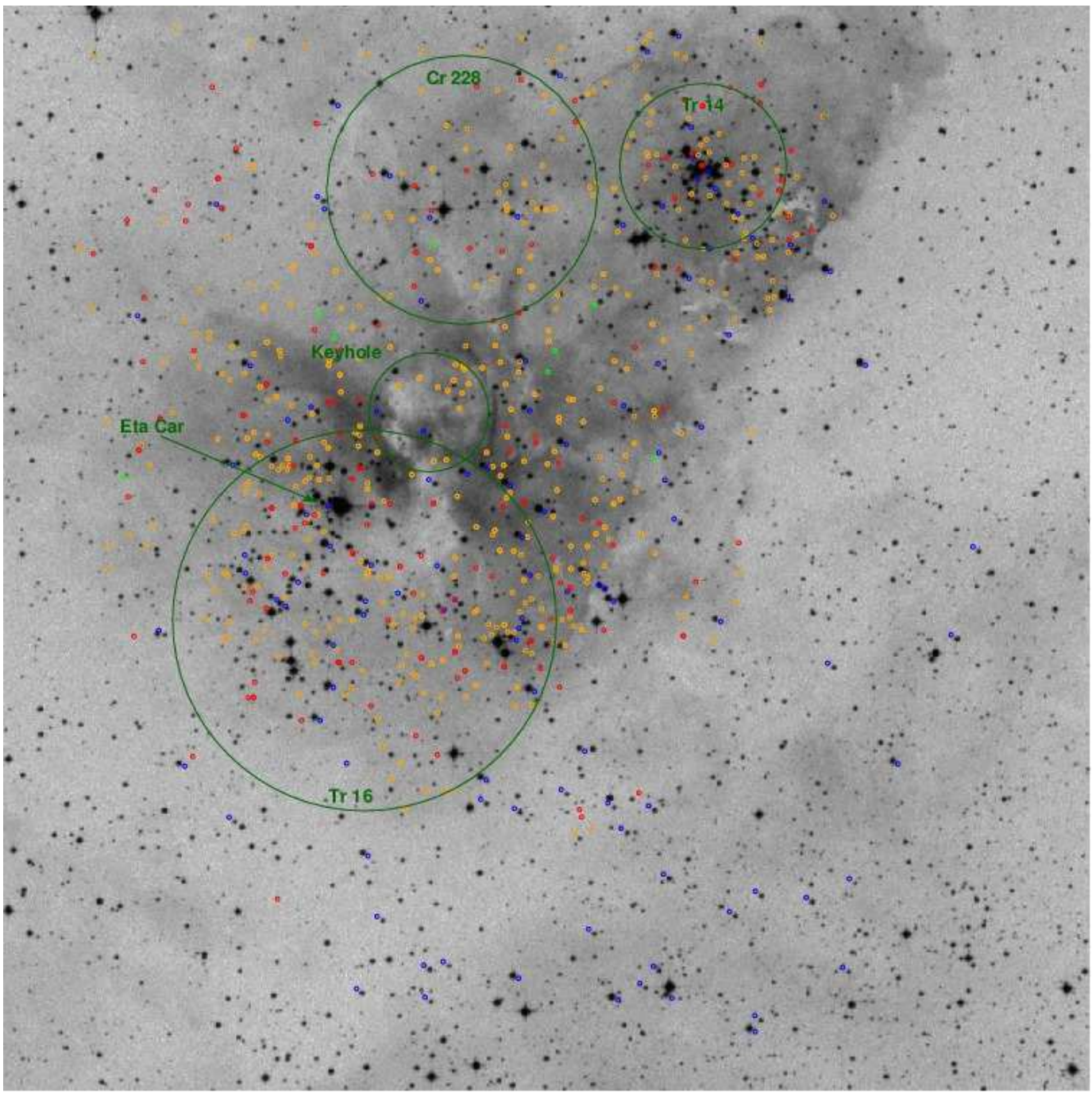}
\caption{DSS1 image of the central part of the Carina Nebula, of size
$26^{\prime} \times 26^{\prime}$, corresponding to 17.4~pc on a side at
the Nebula distance of 2.3~kpc. North is up and East is to the left.
Red circles indicate
positions of sky fibres, using Giraffe setup HR15N. Orange circles
indicate fibres on faint stars with HR15N. Blue circles are sky fibres
using setup HR14A, and light-green circles are UVES sky fibres.
Big dark-green circles indicate the approximate positions of clusters
studied here, and the Keyhole nebula. The dark-green arrow indicates the
star $\eta$~Car.
\label{dss}}
\end{figure*}

The spectra used in this work were obtained mostly with the Giraffe
HR15N setup ($R \sim 19000$), in the wavelength range 6444-6818\AA.
Thirteen OBs were executed using this setup in Carina, yielding 185 sky
spectra, at 137 individual sky positions (35 sky positions were observed
more than once). Using Giraffe setup HR14A, 114
sky spectra are also available, in the wavelength range 6301-6689\AA\ 
($R \sim 18000$),
overlapping that of setup HR15N\footnote{All targets observed using
setup HR14A were also observed with Giraffe setups HR3 (4033-4201
\AA), HR5A (4340-4587 \AA), and HR6 (4538-4759 \AA),
which however do no show strong nebular lines and are therefore not
considered here.}.
Moreover, 15 sky spectra were also
obtained with UVES (580nm setup, $R \sim 47000$),
in the ranges 4768-5802\AA\ (lower arm), and
5822-6831\AA\ (upper arm). The Survey targets are low-mass stars ranging down
to magnitude $V \sim 18.5$: for about half of the 1085 stars observed
with setup HR15N,
the sky brightness at H$\alpha$ (as recorded by the 1.2-arcsec-wide Giraffe fibres)
is 100-1000 times the stellar continuum,
so that high-quality information on the sky emission, at least in the
H$\alpha$ core, was obtained also from the spectra of these faint
stars\footnote{The typical peak H$\alpha$ brightness in classical T~Tauri stars
rarely (if ever) exceeds 10 times the star continuum level, see e.g.\ Cohen and
Kuhi (1979). Moreover, most member stars in Carina are expected to be
weak-line T~Tauri from their weak near-IR excesses (Albacete-Colombo et
al.\
2008), and therefore have still weaker H$\alpha$ emission.}.
In total, we have some information for more than 650 sky positions across the
Nebula. All spectra presented here are from internal data release {\em GESiDR4}.

Figure~\ref{dss} shows a DSS image of the Nebula, with indicated
the sky positions studied here. The bright diffuse emission is bounded
towards South by a 'V' shaped dust obscuration. $\eta$ Car is the
brightest star, a few arcmin East from image center, and the cluster
Trumpler~16 is found all around it, projected against the brightest part
of the Nebula. To the North-west is visible the compact, massive cluster
Trumpler~14 (Tr~14). Just West of $\eta$ Car, the obscured region with a
peculiar shape is the so-called 'Keyhole' nebula, already described by
John Herschel. As the Figure shows, our observations cover widely
different regions of the Nebula.

\begin{figure}
\resizebox{\hsize}{!}{
\includegraphics[bb=5 10 485 475]{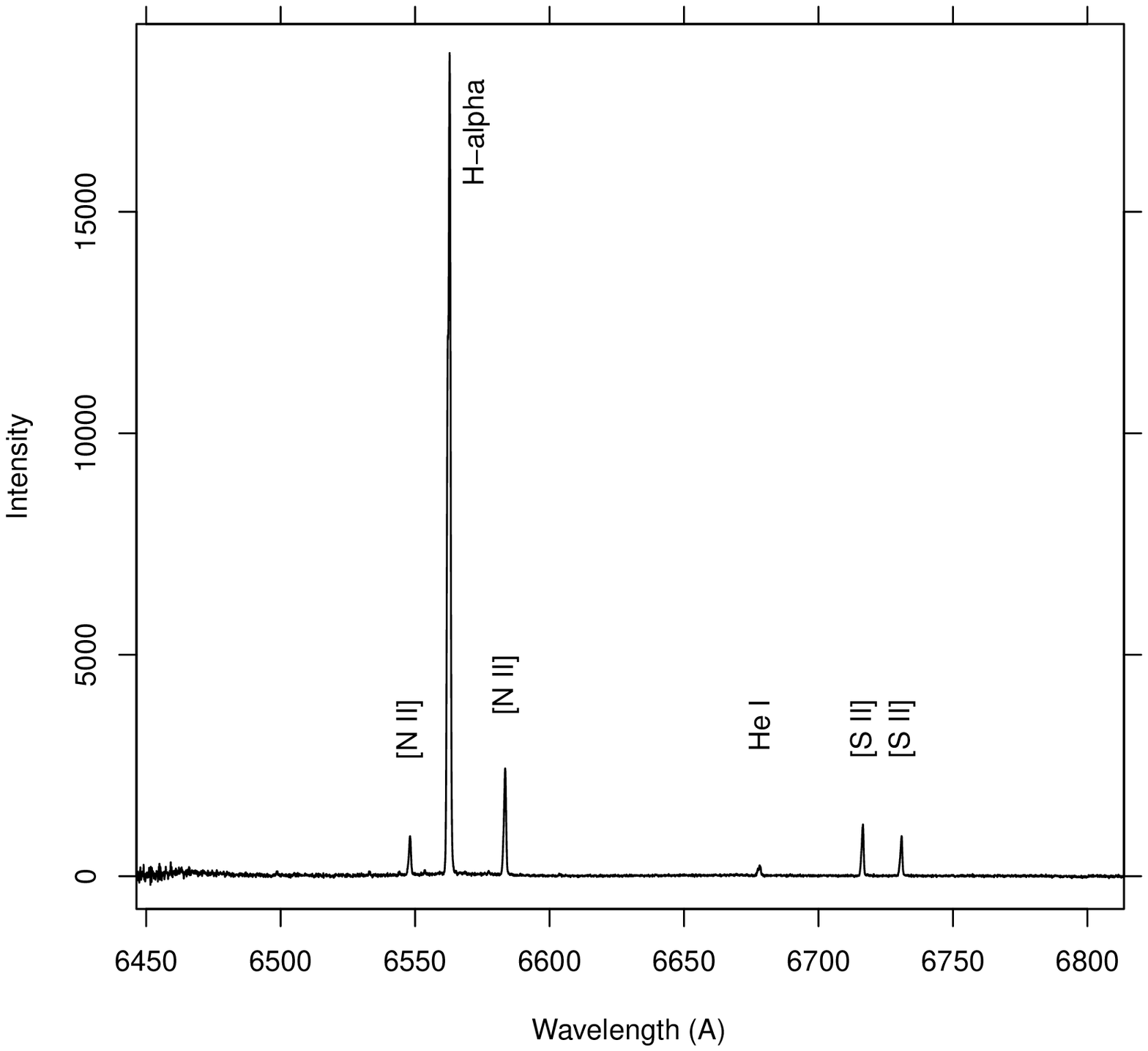}}
\caption{Example of typical sky spectrum in Carina, obtained with
the Giraffe HR15N setup. Prominent nebular lines are labelled.
\label{atlas-sky}}
\end{figure}

The HR15N wavelength range contains several important nebular lines:
H$\alpha$, the neighboring [N II] lines at 6548, 6584\AA, the He~I line at
6678\AA, and the two [S II] lines at 6716, 6731\AA. The HR14A range
contains the same lines except for the [S II] lines. UVES spectra, in
addition to these (and many other) lines, also allow the study of H$\beta$.

\section{Results}
\label{results}

\begin{figure*}
\includegraphics[bb=65 18 544 774,angle=270,width=18cm]{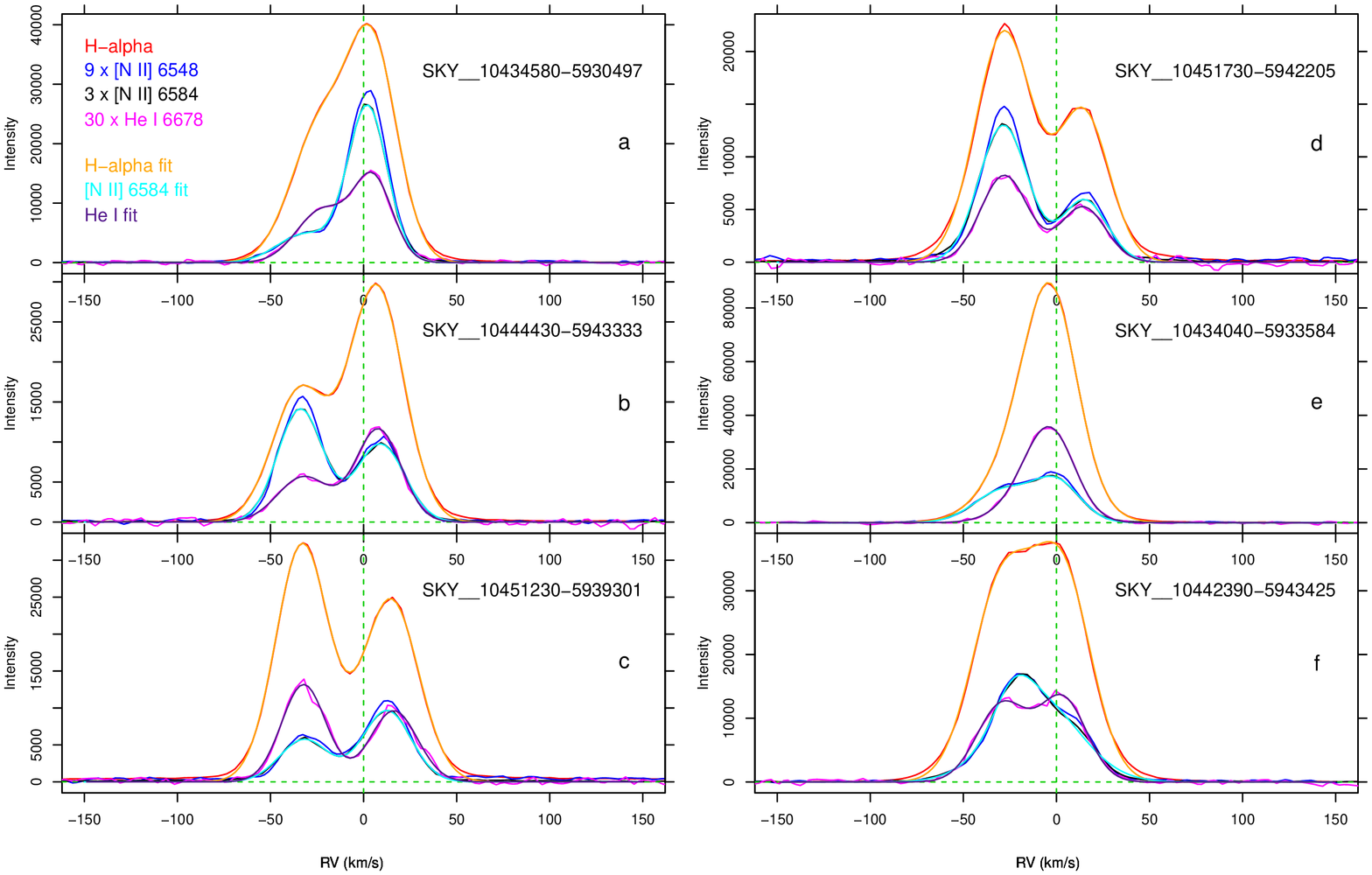}
\caption{Examples of line profiles at H$\alpha$ (red), [N II] 6548 (blue) and 6584\AA
(black), and He~I (magenta), for various positions across the Nebula.
Also shown are two-gaussian best fits for H$\alpha$ (orange), [N II] 6584\AA\
(cyan), and He~I (purple). Note that the lines shown are scaled by the
indicated factors, for ease of comparison with H$\alpha$.
Indicated radial velocities are heliocentric.
\label{atlas-fits}}
\end{figure*}

An example of HR15N sky spectrum is shown in Figure~\ref{atlas-sky}.
From a close inspection, it is seen that most line profiles show
double peaks, as shown in the examples of Figure~\ref{atlas-fits}.
This is not a new discovery, since it was already noted by
several authors
(Deharveng and Maucherat 1975; Walborn and Hesser 1975;
Meaburn, Lopez and Keir 1984),
but the spectral resolution and signal-to-noise
ratio (S/N) offered by our new data permit a deeper study of the
phenomenon. Figure~\ref{atlas-fits} illustrates conveniently the wide
range of emission line profiles found in the Nebula: in H$\alpha$ (red
curves), only panels $b$, $c$, and $d$ show a distinct double-peak
structure; panels $a$ and $f$ show asymmetries in the line, while a
superficial look at panel $e$ does not reveal more than a single
gaussian component. In the latter line profiles of panels $a$, $f$
and $e$, however, two components are clearly seen in the He~I (magenta)
and [N II] lines (blue and black). Looking at the radial velocity (RV)
of the line peaks we observe an approximate correspondence between the
peaks of the different lines. As panels $b$, $c$, and $f$ show clearly,
there is instead no correlation between the intensities of H$\alpha$ and the
corresponding peak in the [N II] line. On the contrary, the intensities of
the He~I peaks show a nearly perfect correspondence with those of
H$\alpha$.

We have fitted all the H$\alpha$, [N II] 6584\AA, and He~I line profiles using
two gaussians; we have not attempted the same fit on the [N II] 6548\AA\
line, since this is weaker by a constant factor 2.95 (set by atomic physics)
than the [N II] 6584\AA\ line. The best-fit functions are also shown in
Fig.~\ref{atlas-fits} with orange (H$\alpha$), cyan ([N II]) and purple (He~I)
colors: the fit to H$\alpha$ is extremely good, being almost
indistinguishable from the observed line profile; even in the case of
distorted, single-peak profiles (e.g., panel $e$) two gaussian
components are actually required to yield a good fit. There are very few
cases where the H$\alpha$ profiles seemed to require a third component, or
even a single component was sufficient: in the vast majority of cases
two gaussians are required. The [N II] and He~I lines are also equally
well fitted by two-gaussian models, although in these cases the
agreement is limited by the lower S/N of the lines (especially He~I).
We present in Tables~\ref{fit-table-1}, \ref{fit-table-2},
and~\ref{fit-table-3} all best-fit parameters of the two-gaussian
models for H$\alpha$, [N II] 6584\AA, He~I, and also [S II] 6717 and 6731\AA\
lines\footnote{Each gaussian component will have the analytical form
$(Norm/\sqrt{2 \pi \sigma^2}) \exp(-(rv-RV)^2/2\sigma^2)$
using the tabulated parameters.}.
In Table~\ref{fit-table-1}, column {\em Type/setup} indicates the
origin of spectrum (sky fibre or faint star), and the Giraffe setup
used; columns $RV$ and $\sigma$ are in units of km/s.

The persistent two components across the whole set of line profiles in
the Nebula (analogous fits were performed on sky-dominated stellar
spectra, and on HR14A spectra, with similar results)
suggests the existence of two well-defined, distinct dynamical components.
Since the gaussian fits represent so faithfully the observed line
profiles, we examine the properties of the two components using the
best-fit parameter values. This allows to study the bulk of nebular
emission in Carina, and will be done in the next Section. A very careful
inspection of Fig.~\ref{atlas-fits}, however, shows that in the far
wings of H$\alpha$ the two-gaussian fit differs from the observed profiles:
these residual emissions are found to possess interesting
characteristics, and will be studied in Section~\ref{resid}.

\subsection{Main components}
\label{main}

\begin{figure}
\resizebox{\hsize}{!}{
\includegraphics[bb=5 10 485 475]{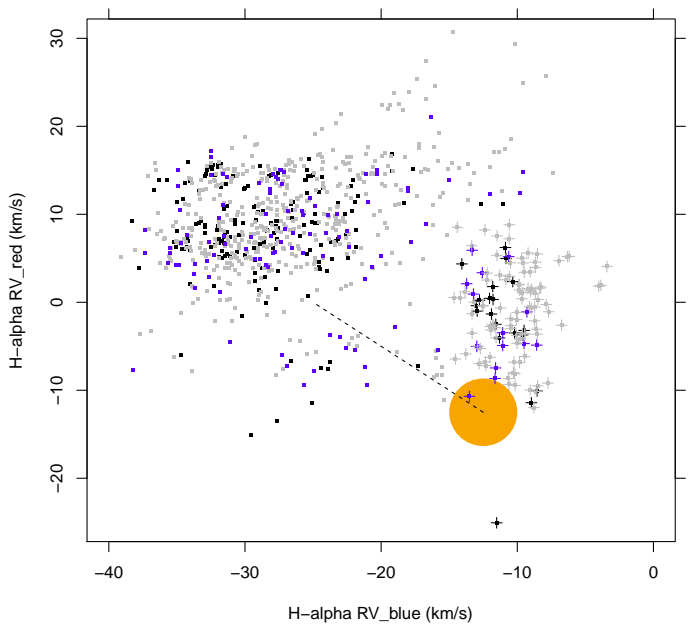}}
\caption{Radial velocity of the receding component vs.\ RV of the
approaching component. Here black dots refer to pure-sky fibres from
HR15N setup, blue dots to pure-sky from HR14A setup, while grey dots to
sky lines from faint stars. The orange circles represent the
average RV of low-mass cluster stars. The dashed line is the expected
locus for purely spherical expansion. A subsample of apparently peculiar
datapoints is highlighted using '$+$' symbols.
\label{ha-rv}}
\end{figure}

From here on, we will refer to the bluer component of the lines as the
'approaching' component, and to the redder one as 'receding', in a
relative sense since they may both happen to have a positive (or
negative) RV on the same sightline.
As the sky image of Fig.~\ref{dss} shows, the bright
nebulosity is centered on the cluster Trumpler~16, so it can be expected
that its dynamics is related to the cluster dynamics.
Wilson (1953) reports for $\eta$ Car a RV of -25~km/s.
We anticipate here that bona-fide
low-mass members of Trumpler~16, from the same Gaia-ESO HR15N dataset
studied here, have RVs in the range of -10 to -15~km/s (Damiani et al., in
preparation), and the low-mass star RVs are usually more accurate than
those from massive stars, because these latter have fewer and broader lines.
We assume therefore a 'fiducial' RV for the Trumpler~16-14 clusters of
-12.5~km/s; in our preliminary analysis, no difference in mean RV is
found between clusters Trumpler~16 and 14.

Figure~\ref{ha-rv} shows the best-fit RVs of the receding H$\alpha$ component
$RV_{red}$ vs.\ that of the approaching H$\alpha$ component $RV_{blue}$.
Datapoints refer to pure-sky HR15N
data (black), sky from faint stars (grey), and pure-sky HR14A data
(blue). The orange circle describes the locus of cluster stars, i.e.,
the center-of-mass velocity of the cluster $RV_{cm}$ (Damiani et al., in
preparation).
It can be seen that datapoints split naturally in two main groups
(equally populated by black, grey and blue dots): a larger group
centered at $(RV_{blue},RV_{red})=(-30,10)$, and a smaller one,
closer to $RV_{cm}$. We have indicated these latter observations
(at $RV_{blue}>-15$ and $RV_{red}<10$) with small plus signs, and we
will refer to these sky locations as ``zero-velocity gas" (with respect to
$RV_{cm}$).
Note that even for zero-velocity datapoints we are still able to fit two
distinct, non-degenerate gaussian components to most lines, thanks to
the very large signal-to-noise ratio of our spectra.
The dashed line starting at $RV_{cm}$ is the expected locus for an
ideal, spherical, optically thin expanding region, whatever its
geometrical thickness or velocity law. The requirement of symmetry, in
fact, makes that for each line of sight towards it two RV values are
observed, symmetrically on opposite sides of $RV_{cm}$ (i.e., the
velocity of the center of expansion). This behaviour is actually shown
by planetary nebulae (Osterbrock and Ferland 2006).
Fig.~\ref{ha-rv} shows instead that no global spherical expansion occurs
in the {studied part of the} Carina Nebula, and
there is no one-to-one correspondence between $RV_{blue}$ and $RV_{red}$
on the same line of sight. Only the average approaching and receding RVs of
the main group of datapoints are found on the prolongement of the dashed
segment: on average, the approaching and receding RVs are respectively equal to
$RV_{cm} -20$ and $RV_{cm} +20$ km/s, but no well-defined center of expansion
is identifiable in the data.

Since the datapoints are not smoothly
connected to the center-of-mass velocity, the approaching and receding
components resemble more two distinct layers of gas than a spherical
distribution; alternatively, the distribution might still be spherical,
but our observations only cover its central parts, and omit parts where
expansion is orthogonal to the line of sight, and thus at $RV \sim 0$.
Smith and Brooks (2007) discuss the large-scale diffuse emission
in Carina, made of several shell-like features roughly centered on
Trumpler~14-16 of sizes 20-30~pc, and suggest that the double-peaked optical
lines would arise from opposite sides of those shells, along the line of sight.
This hypothesis, however, fails to
explain why the intensity of the H$\alpha$ emission is so spatially
concentrated near the central clusters, at intensities much larger than
the outer border of the shells, while projection effects would cause the
reverse effect (largest dilution at the center). The line-of-sight depth of the
regions producing the most intense H$\alpha$ emission can be expected to be
of order of its sky-projected size, or $\sim 10$~pc in diameter.
This might be a sufficiently large size to account for
the lack of precise correspondence between $RV_{blue}$ and 
$RV_{red}$ along the same line of sight.
We must not forget, nevertheless, the known complexity of the Carina
SFR, and the possibility that there are actually several distinct
centers of expansion, perhaps at different RVs. The observed emission
might therefore be a chaotic superposition of different expanding spheres.

\begin{figure}
\resizebox{\hsize}{!}{
\includegraphics[bb=5 10 485 475]{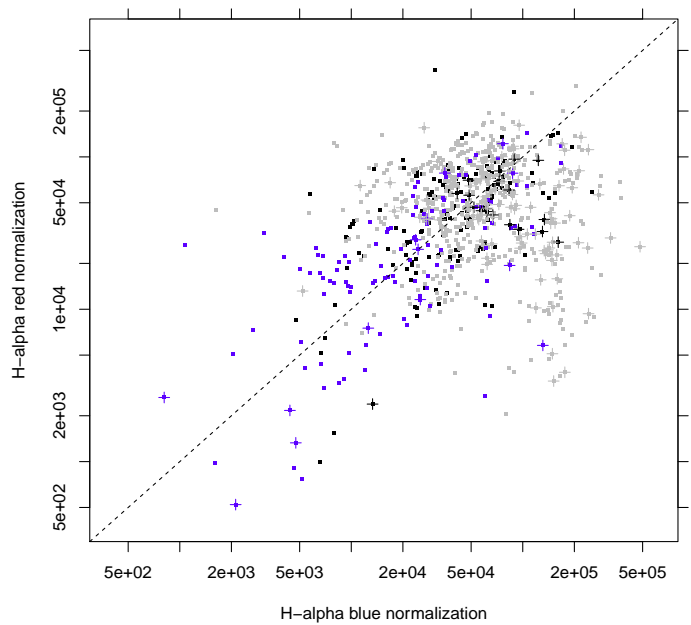}}
\caption{Normalizations of best-fit gaussians for the receding vs.\
approaching components. The dashed line represents identity. Other
symbols as in Fig.~\ref{ha-rv}.
\label{ha-norm}}
\end{figure}

Figure~\ref{ha-norm} shows the intensity of H$\alpha$ of the approaching component
vs.\ that of the receding component. Only a weak correlation exists
between the two, dominated by the datapoints (blue dots) from the HR14A
dataset, coming partly from stars outside the strong nebulosity, to the
South-west. The intensity pattern in the receding component is therefore
sensibly different in its morphological details than that of the
approaching component, a result that H$\alpha$ images might have never shown.
The zero-velocity gas already defined in Fig.~\ref{ha-rv} is also indicated
in Fig.~\ref{ha-norm} with small plus signs.
We also observe
that the intensity of the H$\alpha$ lines from the receding and approaching gas
are on average similar: if they have intrinsically similar brightnesses,
then the absorption towards them cannot be sensibly different (much less
than one magnitude of extinction), and only a low amount of dust may
exist between them, despite the large distance which is
likely to separate them.

\begin{figure}
\resizebox{\hsize}{!}{
\includegraphics[bb=5 10 485 475]{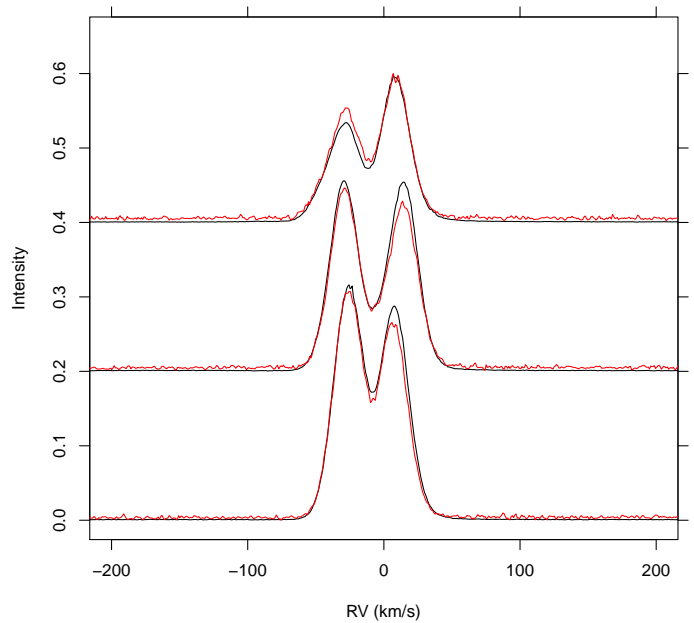}}
\caption{Examples of H$\beta$ (red) and H$\alpha$ (black) lines from pure-sky
fibres using UVES. A vertical shift was applied to spectra from different sky
positions. H$\beta$ lines were scaled up by a factor of 7.
\label{uves-hbeta}}
\end{figure}

Thus far we have assumed implicitly that the receding gas is generated
from expansion starting from some place inside clusters Trumpler~16
and~14, and is located at larger distances than the clusters themselves,
while the opposite holds for the blue-shifted, approaching gas. In
principle, it would not be impossible to have the reversed situation, where the
two layers are actually moving towards one another, leading to a huge
collision, which might have already generated the existing massive
clusters. In order to discriminate between the two scenarios, it is
necessary to determine unambiguously which layer is closer to us. We
have seen that extinction differences, if they exist, are subtle, and
cannot be determined from the statistical arguments discussed above.
However, a much more accurate way to establish the reddening suffered by
optical line emission is through the Balmer decrement. For this aim we
use the 15 UVES pure-sky spectra, which include both H$\alpha$ and
H$\beta$. Being no spectrophotometric instrument, we cannot determine
absolute reddenings for the two layers (i.e., calibrated flux ratios
between H$\alpha$ and H$\beta$); however, it is possible to find
which layer has the largest reddening of the two. Figure~\ref{uves-hbeta}
shows examples of comparison between H$\alpha$ (black) and H$\beta$ (red): in
all cases the receding, red-shifted layer shows slightly attenuated H$\beta$
intensity with respect to H$\alpha$, and thus larger reddening. The reverse is
never found. This may be taken as evidence that the two layers are
expanding away from the central clusters and $\eta$ Car, although a
different temperature between the two layers might mimic the same effect
(Osterbrock and Ferland 2006).

\begin{figure}
\resizebox{\hsize}{!}{
\includegraphics[bb=5 10 485 475]{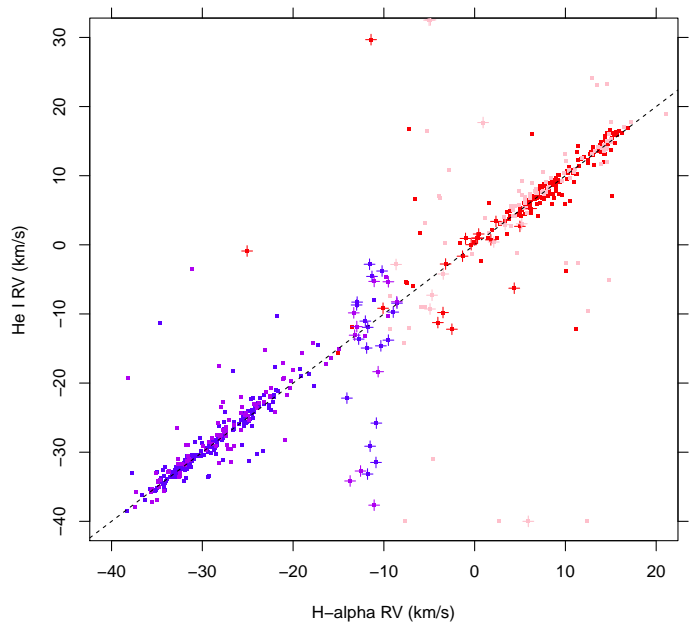}}
\caption{Radial velocity from the He~I 6678~\AA\ line vs.\ RV from
H$\alpha$.
Red (blue) symbols refer to the receding (approaching) component from
HR15N data. Purple (pink) symbols refer to the same two components, but
from HR14A pure-sky fibres.
Plus symbols refer to ``zero-velocity" gas as in previous Figures.
The dashed line represents identity.
\label{ha-he-rv}}
\end{figure}

\begin{figure}
\resizebox{\hsize}{!}{
\includegraphics[bb=5 10 485 475]{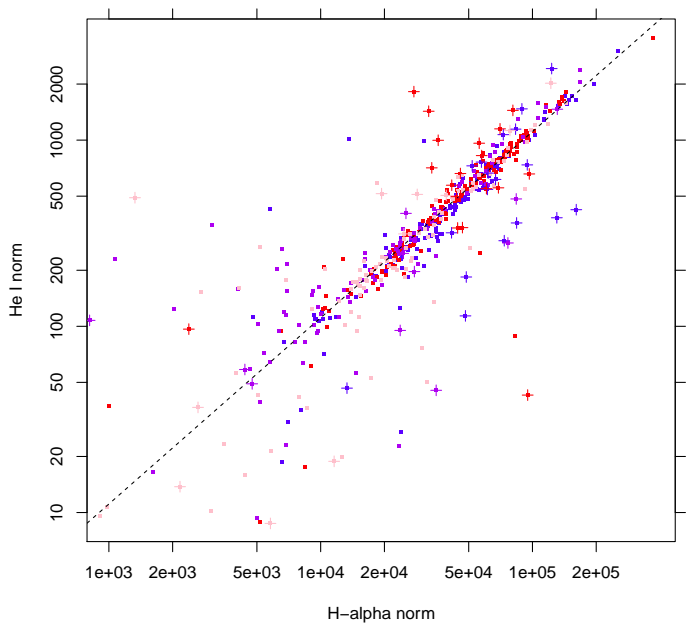}}
\caption{Fit normalization for He~I vs.\ normalization for H$\alpha$.
Symbols as in Fig.~\ref{ha-he-rv}.
The dashed line indicates a constant ratio of 1/90.
\label{ha-he-norm}}
\end{figure}

We have compared best-fit parameters between different lines, starting
from H$\alpha$ and He~I. The respective RVs are shown in
Figure~\ref{ha-he-rv}, where red (blue) dots indicate the receding
(approaching) components from HR15N data, and pink (purple) dots the
same components from HR14A data, respectively. The agreement is
excellent, except for a few outliers, most of which already known as the
``zero-velocity" gas (plus signs). It should be noted that the He~I line
is by far the weakest line in our sample (Fig.~\ref{atlas-sky}), with
intensity $\sim 1/90$ of H$\alpha$, so that in Fig.~\ref{ha-he-rv} errors in
the ordinates must be much larger than errors in the abscissae; the latter
are expected to be of the same size as the plotted dots. The very good
coincidence in RVs suggests strongly that the same gas is emitting
in these two lines. This conclusion is reinforced
by the comparison of the intensities in the two lines in
Figure~\ref{ha-he-norm}, which shows again a very tight correlation.

\begin{figure}
\resizebox{\hsize}{!}{
\includegraphics[bb=5 10 485 475]{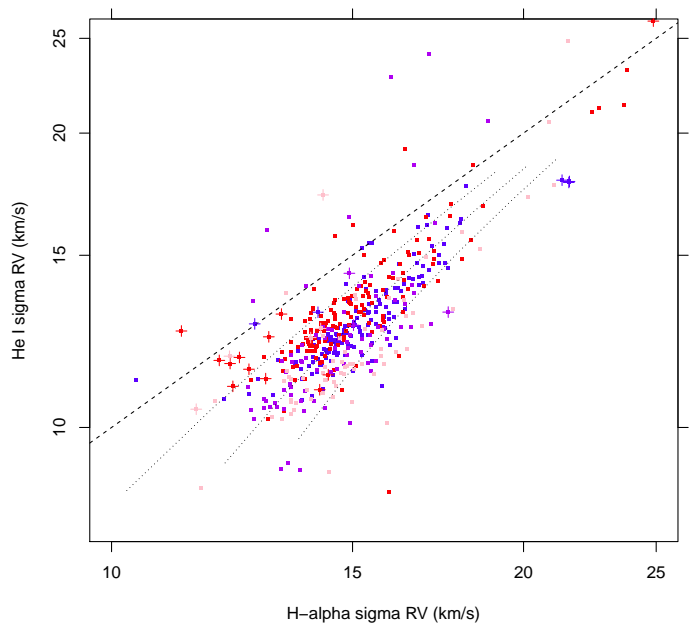}}
\caption{Gaussian widths $\sigma$ of He~I lines vs.\ those of H$\alpha$.
The three dotted lines represent loci of constant gas temperature,
$T_{gas}=5000,10000,15000$ from top to bottom.
The dashed line represents equality.
Symbols as in Fig.~\ref{ha-he-rv}.
Only datapoints with the same RV from H$\alpha$ and He~I within 3~km/s are shown.
\label{ha-he-sigma}}
\end{figure}

Since the same material emits in H$\alpha$ and He~I lines, the slight
difference in their line widths, already observed with reference to
Fig.~\ref{atlas-fits} above, must be related to microscopic properties of
the gas, and can be taken as a measure of temperature. The total
line width for hydrogen will be:
\begin{equation}
\sigma_H = \sqrt{\sigma_T^2 + kT/m_H}
\end{equation}
and an analogous expression, with the appropriate atomic mass, for He.
$\sigma_T$ is the component of the line width due to instrumental width and
turbulence (and any other macroscopic velocity field,
identical for hydrogen and helium, and therefore irrelevant for our
derivation of temperature).
Therefore, we plot in Figure~\ref{ha-he-sigma} the H$\alpha$ and He~I line
widths, with three loci of constant temperature (5000, 10000, 15000~K).
Most datapoints fall between 5000-15000~K; strong outliers are more likely to
arise from failed fits than from extremely high (low) temperatures.
Datapoints for the approaching (blue) gas tend to show larger
temperatures than those of receding gas, by a few thousands K.
Turbulent velocities span a range of approximately 7-15~km/s (net of the
instrumental line width), often above the sound speed for $T\sim 10000$~K.

\begin{figure}
\resizebox{\hsize}{!}{
\includegraphics[bb=5 10 485 475]{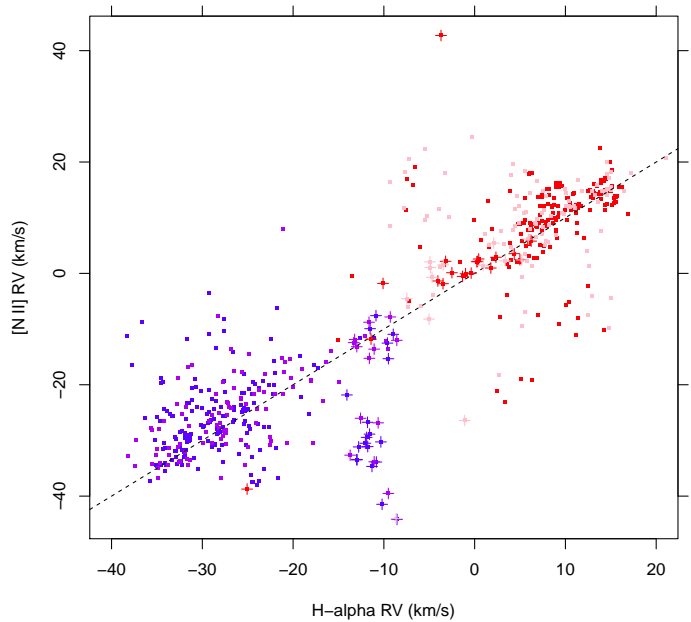}}
\caption{Radial velocity from the [N II] 6584~\AA\ line vs.\ RV from
H$\alpha$.
Symbols as in Fig.~\ref{ha-he-rv}.
\label{ha-n2-rv}}
\end{figure}

Next, we have compared H$\alpha$ parameters with those for [N II] 6584\AA\
(the strongest of the [N II] doublet). The RV comparison is shown in
Figure~\ref{ha-n2-rv}: the agreement between the respective RVs is much
worse than in the case of He~I of Fig.~\ref{ha-he-rv}. The [N II] line
is about 20 times stronger than the He~I line, so the disagreement is
certainly not due to errors, but is real, and a definite indication that
the gas emitting H$\alpha$ is not strictly the same as that emitting the [N II]
lines. Although not the same gas, the dynamics of these two components
is similar, since the H$\alpha$ and [N II] RVs are nevertheless well
correlated (the receding gas more so than the approaching gas): they
might be two adjacent layers or bubbles in the same expanding material.
We note that there is no definite sense in which the respective RVs
differ: if for example the H$\alpha$-emitting gas were moving systematically
faster than the [N II]-emitting gas, one would observe a correlation
with a slope different from 1 (the dashed line), still centered on
$RV_{cm}$. The ``zero-velocity" outliers (plus signs) are found in [N II]
sometimes at $RV \sim RV_{cm}$, and sometimes at definite non-zero RV.
We will return to this point below.

\begin{figure}
\resizebox{\hsize}{!}{
\includegraphics[bb=5 10 485 475]{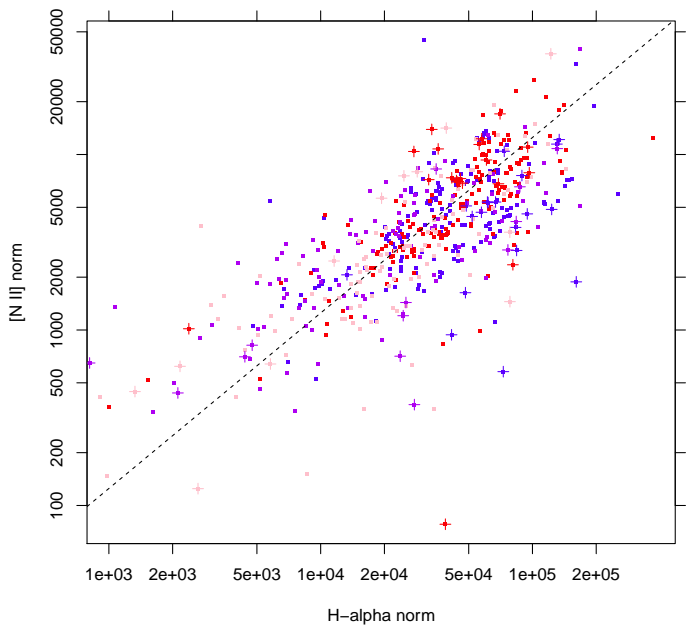}}
\caption{Fit normalization for [N II] vs.\ normalization for H$\alpha$.
The dashed line indicates a constant ratio of 1/8.
Other symbols as in Fig.~\ref{ha-n2-rv}.
\label{ha-n2-norm}}
\end{figure}

\begin{figure}
\resizebox{\hsize}{!}{
\includegraphics[bb=5 10 485 475]{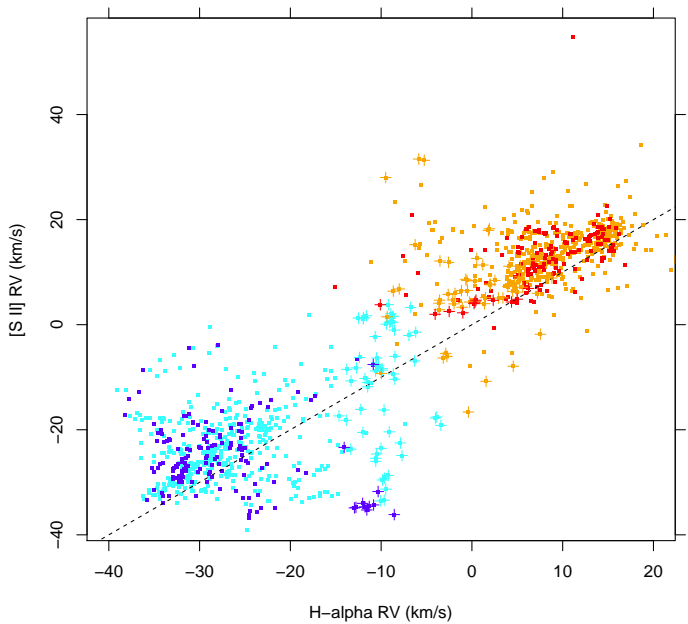}}
\caption{Radial velocity from the [S II] 6717~\AA\ line vs.\ RV from
H$\alpha$.
Red (blue) symbols refer to the receding (approaching) component from
HR15N data. Cyan (orange) symbols refer to the same two components, but
from HR15N observations of faint stars.
Other symbols as in Fig.~\ref{ha-he-rv}.
\label{ha-s2-rv}}
\end{figure}

The comparison between the intensities of H$\alpha$ and [N II] strenghtens
the conclusion that they arise from different material even more
(Figure~\ref{ha-n2-norm}), with intensity ratios between the two lines
spanning an order of magnitude (or more). We will discuss below the
interpretation of this fact.

\begin{figure}
\resizebox{\hsize}{!}{
\includegraphics[bb=5 10 485 475]{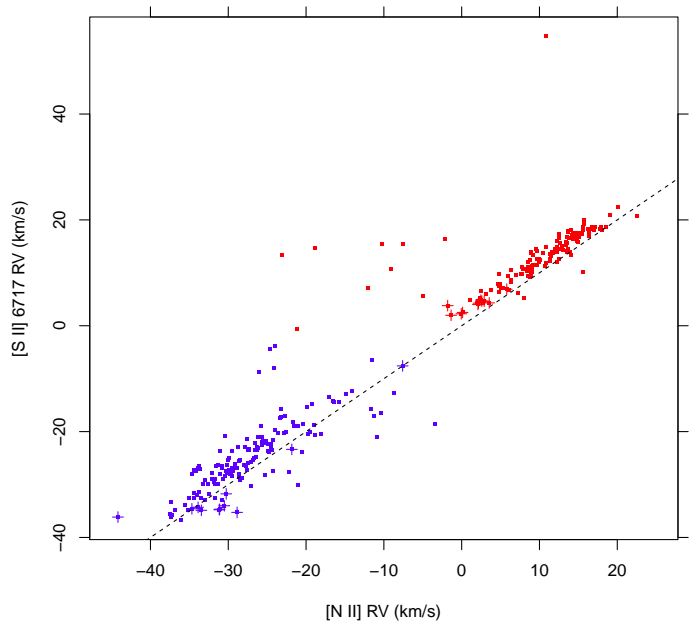}}
\caption{Radial velocity from the [S II] 6717~\AA\ line vs.\ RV from [N
II] 6584\AA.  Symbols as in Fig.~\ref{ha-he-rv}.
\label{n2-s2-rv}}
\end{figure}

\begin{figure}
\resizebox{\hsize}{!}{
\includegraphics[bb=5 10 485 475]{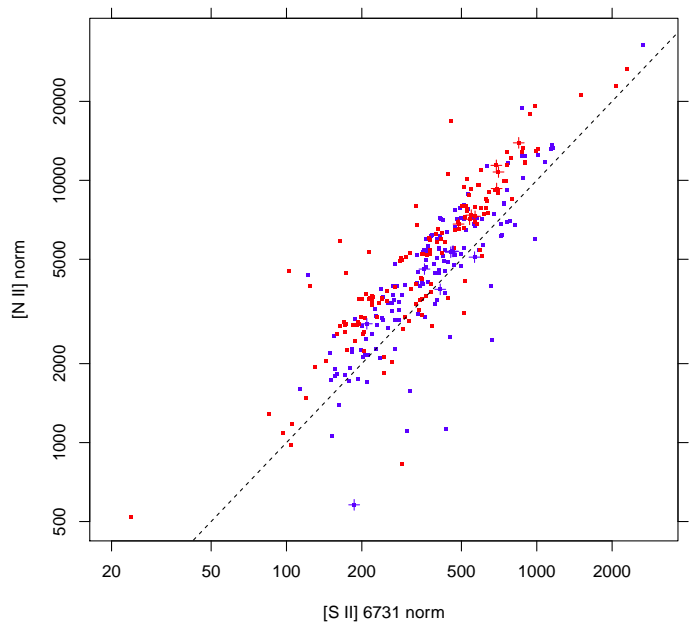}}
\caption{Fit normalization of [N II] 6584\AA\ vs.\ [S II] 6731~\AA.
Symbols as in Fig.~\ref{ha-he-rv}.
\label{s2-n2-norm}}
\end{figure}

The comparison between the RVs from H$\alpha$ with those from the [S II]
6717\AA\ line is shown in Figure~\ref{ha-s2-rv}. Here, many more
datapoints are shown than in Fig.~\ref{ha-n2-rv}, from observations of
sky lines from faint-star HR15N spectra. On the other hand, datapoints
from HR14A observations are missing since the [S II] lines are outside
their wavelength range. Fig.~\ref{ha-s2-rv} shows the same pattern as
Fig.~\ref{ha-n2-rv}, with a loose correlation and a peculiar place for
the zero-velocity gas. This suggests at least a similarity in the
dynamics of the [S II]-emitting and [N II]-emitting gas, which is
confirmed by their direct comparison, shown in Figure~\ref{n2-s2-rv}. Despite
the fact that both lines are stronger than He~I, the latter correlation is not
as tight as that between H$\alpha$ and He~I, so that again the conclusion is
that the [S II]-emitting gas is nearly coincident, but not identical,
to the [N II]-emitting gas. In Fig.~\ref{n2-s2-rv} we also note that
most of zero-velocity datapoints are not outliers here, but follow the
same correlation as other datapoints. Since the plotted data are
statistically independent, this demonstrates that those points do not
arise from failed fits, but have a real significance.
The same considerations apply to the comparison between [S II] and [N
II] intensities in Figure~\ref{s2-n2-norm}: a good correlation, but
significantly different from identity.

\begin{figure}
\resizebox{\hsize}{!}{
\includegraphics[bb=5 10 485 475]{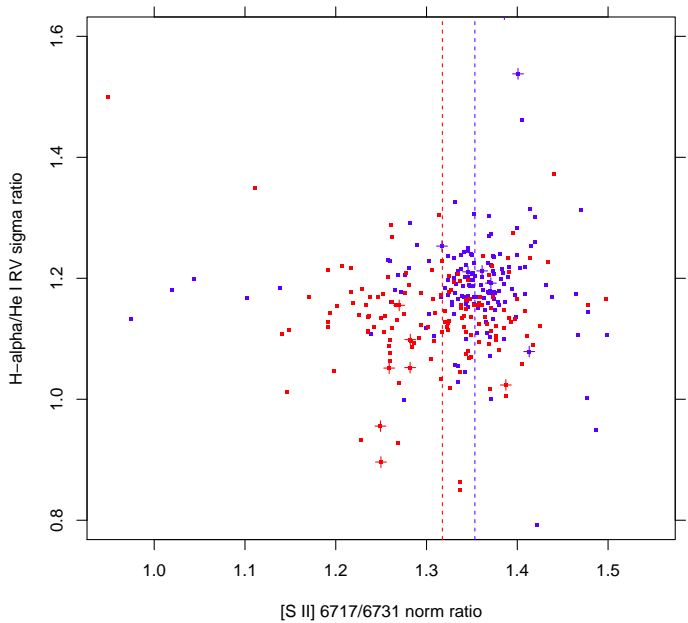}}
\caption{Gaussian width ratio between H$\alpha$ and He~I (proxy for
temperature), vs.\ normalization ratio between [S II] 6717/6731\AA\ (proxy
for density). Temperature increases upwards, while density increases
towards the left.  Symbols are as in Fig.~\ref{ha-he-rv}.
The red (blue) vertical dashed line indicates the median [S II]
6717/6731\AA\ ratio for the receding (approaching) components. On average,
the receding component has higher density and lower temperature than the
approaching component.
\label{s2-ha-he-norm}}
\end{figure}

Following this preliminary assessment, we have examined the ratio
between the two [S II] 6717/6731\AA\ lines, which depends only on the gas
density, for densities in the range $10^2-10^5$ cm$^{-3}$ (Osterbrock
and Ferland 2006). A plot of the ratio between the H$\alpha$ and He~I line widths (a
proxy for temperature as discussed above) vs.\ the [S II] 6717/6731\AA\
intensity ratio is shown in Figure~\ref{s2-ha-he-norm}.
There is a tendency for the approaching gas to have both higher
temperatures and lower densities (higher 6717/6731\AA\ ratio) than the
receding gas; median 6717/6731\AA\ ratios are indicated by the vertical
dashed lines, and correspond to densities of 100 ($\times \sqrt{10^4/T}$)
cm$^{-3}$, and 200 ($\times \sqrt{10^4/T}$) cm$^{-3}$, respectively.

\begin{figure}
\resizebox{\hsize}{!}{
\includegraphics[bb=5 10 485 475]{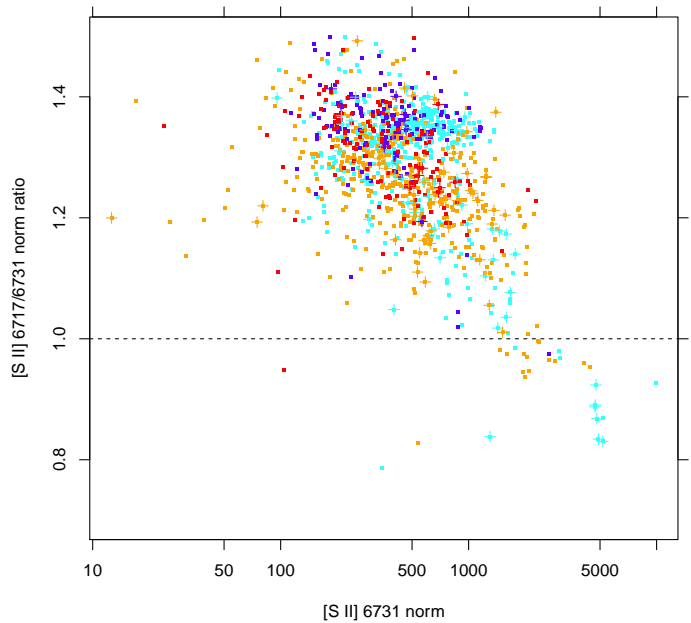}}
\caption{Intensity ratio between [S II] 6717\AA\ and 6731\AA\ lines,
vs.\ intensity of [S II] 6731\AA.
Symbols as in Fig.~\ref{ha-s2-rv}.
The dashed line at unity ratio corresponds to electron
density $N_e \sim 600$ cm$^{-3}$.
\label{s2-norm-ratio}}
\end{figure}

A plot of the [S II] 6717/6731\AA\ intensity ratio vs.\ intensity of the [S
II] 6731\AA\ line is instead shown in Figure~\ref{s2-norm-ratio}.
The correlation, although not strict, indicates that for increasingly
strong [S II] emission the density of the emitting gas also increases,
with the largest electron densities found being $\sim 2 \times 10^3$
($\times \sqrt{10^4/T}$) cm$^{-3}$.

\subsection{Low intensity components}
\label{resid}

\begin{figure*}
\includegraphics[bb=30 58 590 754,width=18cm]{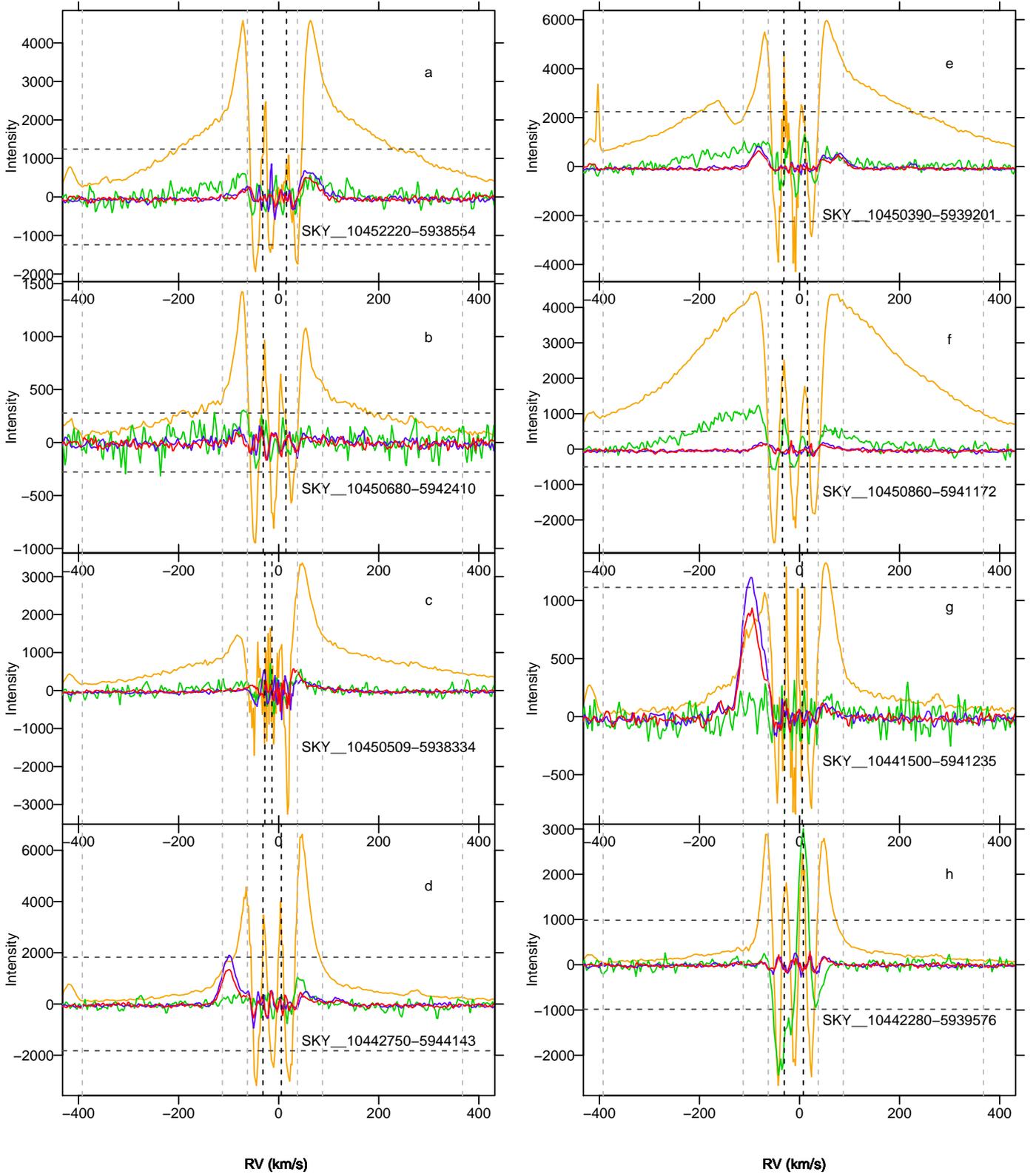}
\caption{Examples of low-intensity components in the lines, as residuals
from the gaussian best fits to the main components of
H$\alpha$ (orange), {He~I 6678 (green)}, [S II] 6717\AA\ (blue) and
6731\AA\ (red). {The He~I residuals are scaled up by a factor of 10}, and
those of [S II] by a factor of 3.
Black vertical dashed lines indicate the RVs of the two main gaussian
components. Grey dashed lines indicate the adopted boundaries to define
the wide and narrow H$\alpha$ wings, all centered at RV=$-12.5$~km/s.
Indicated radial velocities are heliocentric.
The emission features in the H$\alpha$ profiles at RV$\sim -420$ and $\sim +280$
km/s originate from terrestrial airglow.
In the plotted units, H$\alpha$ peak emission before best-fit subtraction is
for each panel respectively equal to 124300 (panel $a$), 27800 ($b$), 426700
($c$), 182600 ($d$), 224200 ($e$), 50200 ($f$), 111300 ($g$), and 98300 ($h$).
{Intensity levels of $\pm 1$\% of peak emission are indicated in each
panel by horizontal dashed lines.}
\label{atlas-resid}}
\end{figure*}

The two emission components well described by gaussian shapes constitute
the bulk of the nebular emission in the studied region. However, as
anticipated from close inspection of Fig.~\ref{atlas-fits}, there are
small but significant departures from these pure gaussian shapes in
their extreme wings. We are able to study these low-intensity wings
thanks to the very high S/N of our data.
To this purpose we consider spectra from sky fibers only, to avoid
any possibility of confusion with stellar features.
After subtracting out the
best-fit gaussian components from the lines, residual emission can in
most cases be seen, of which representative examples are shown in
Figure~\ref{atlas-resid}. The intensity levels of this emission are
about two orders of magnitude lower than the peak H$\alpha$ emission.
Figure~\ref{atlas-resid} shows residuals in H$\alpha$ (orange), {He~I
6678 (green),} and the [S II]
lines (6717\AA\ in blue, and 6731\AA\ in red).
The two black dashed lines indicate the best-fit RVs
of the main gaussian approaching/receding components. The grey dashed
lines indicate RVs of $RV_{cm} \pm (50,100,380)$ km/s, for
reference\footnote{The 380~km/s boundary was chosen so that the
measurement of wide-wing intensity below remains unaffected by the skyglow
feature falling at $RV \sim -420$~km/s.}.
{We demonstrate in the Appendix that the residual patterns seen in
Fig.~\ref{atlas-resid} are not of instrumental origin.  }
First, we note from the Figure that the residual emission in H$\alpha$ shows
two apparently distinct components: a lower-intensity high-velocity gas,
with RV reaching (absolute) values up to 400~km/s or above; and a narrow
component, mostly seen at lower absolute RVs between 50-100~km/s (relative to
$RV_{cm}$), i.e.\ between the two bands delimited by grey lines near the
center. In [S II] (red/blue curves) no wide wings are seen, but the
narrow components may sometimes appear, at RVs close to those of
H$\alpha$
narrow wings.
A wide variety of situations is encountered across the Nebula: both the
wide and narrow wings span a wide range of intensities, and each type
may be observed without the other (see e.g., panels $f$ and $h$).
The wide wings present a marked symmetry between the blue and red sides,
while the narrow wings much less so: the red wings may be stronger than
the blue one, or viceversa (panels $b$ and $c$).
When the [S II] emission is present, although it resembles the emission in
the narrow H$\alpha$ wings, it may be found at absolute RVs both lower (panel $a$)
and higher (panels $d$,$g$) than this latter; in panel $g$ it is very strong
in relative terms.
Finally, panel $e$ shows a conspicuous case (not the only present) of
self-absorption in the inner portion of the blue wide wing, at about
-100~km/s; the width of this absorption component is considerable, being
$\sim 50$ km/s.
In this section we try to understand the physical meaning of all these
features.

In the Figure, one may also note a regular saw-tooth pattern in the
residuals near RV=0: this is not random noise, but a systematic effect.
The maximum amplitude of such residual pattern, compared to original line peaks
(see Figure caption), is of order of 1-2\%, in accordance with the very
good appearance of fitted models in Fig.~\ref{atlas-fits}.
From numerical experiments, it turns out that this effect arises because
no wide emission component was included in our fitting model, in
addition to the two gaussian components: as a result, the widths
$\sigma$ of the best-fitting gaussians are very slightly overestimated
(by $\sim 1.5$\%), producing exactly the same residual pattern as
observed.
This affects marginally only the immediate
neighborhood of the main components, and therefore also the inner
boundary of the narrow H$\alpha$ components; further out, however, the
residual line profiles are totally unaffected.
Incidentally, if we are able to detect systematic effects in the
best-fit widths $\sigma$ of such a small amplitude, random errors on
$\sigma$ are necessarily much smaller than 1\% on average, since
otherwise they would mask the above effect.
{In general, even where wide wings are not detected, 
additional components of low-amplitude, compared to the main ones, will
produce similar residual patterns, especially in H$\alpha$.
Therefore, we consider questionable the physical existence of gas
emitting in the narrow H$\alpha$
wings. Instead, the emission features in the [S II] residuals between
(absolute) velocities 50-100 km/s, of amplitude much larger than the [S
II] residuals in the range [$RV_{cm}-50,RV_{cm}+50$] km/s,
can be considered as a real phenomenon.}

We remark that Walborn et al.\ (2002) have found many RV components (in
absorption) in the STIS UV spectra of four O stars in Carina, in lines
of several ionized species: for most lines, however, the STIS spectra
are saturated in RV range
studied in the present paper, and provide no useful information,
while the many narrow absorption components at RV$<-100$~km/s found in the
STIS spectra have a
completely different appearance than the high-velocity components studied
here, and therefore originate in {different regions.}

The existence of the wide H$\alpha$ wings was already known for decades
(Lopez and Meaburn 1984); in the direction of the Keyhole nebula,
Walborn and Liller (1977) and Boumis et al.\ (1998) found that they are the
spectrum of $\eta$ Car, reflected from local dust.
{This explains the apparent self-absorption at $RV \sim -100$ km/s,
seen in Fig.~\ref{atlas-resid} (panel $e$), and in a few other places
near to that position.
We will discuss in the next section the spatial distribution of the wide-wing
emission, which is strongest in coincidence with the Keyhole and along the line
between it with $\eta$ Car, as found by Boumis et al.\ (1998).
This again suggests reflected emission from $\eta$ Car for their origin,
rather than local high-velocity gas.
Our data enable us to test further these two alternative hypotheses.
A detailed comparison with the actual spectra of $\eta$~Car can be done
using data from the ESO archive, where 16 UVES spectra are available
covering the years 2002-2004; of them we only consider those (eight) without
strong signs of saturation in H$\alpha$. A representative selection is shown
in Figure~\ref{eta-car-spectra}, in the regions around H$\alpha$ and He~I
lines. We looked for similarities between our residual patterns of
Fig.~\ref{atlas-resid} and
the $\eta$~Car spectra in the HR15N wavelength range. The most
characteristic features seen in the $\eta$~Car spectra are the wide
H$\alpha$
wings and a peculiarly-shaped He~I 6678 line, as seen in
Fig.~\ref{eta-car-spectra}. Like H$\alpha$, also the profile of the $\eta$~Car
He~I 6678 line is variable.  We find a striking similarity between the latter
line profiles and our He~I fit residuals of Fig.~\ref{atlas-resid},
panels $a$, $e$, and $f$, and some hint of blue-wing He~I residual
emission also in panels $b$ and $g$. This would support the case for
wide H$\alpha$ wings as a reflection effect even more. }
{We will show below that there is no proportionality between the
sky continuum intensity (from reflection nebulosity) and the wide-wing
intensity. This can be reconciled with wide wings being due to
reflection of $\eta$~Car emission only if the sky continuum is larger
than the scattered $\eta$~Car continuum, due to other stellar continua
contributing as well. We find that this is indeed the case:
Figure~\ref{wings-continuum-etacar} shows that the $\eta$~Car continuum
and wide wings (downscaled by a suitable factor) describe well the lower
envelope of nebular datapoints. Therefore, the missing correlation
between the two quantities across the nebula is not inconsistent with
the reflection hypothesis for the wide wings.

\begin{figure}
\resizebox{\hsize}{!}{
\includegraphics[bb=5 10 485 475]{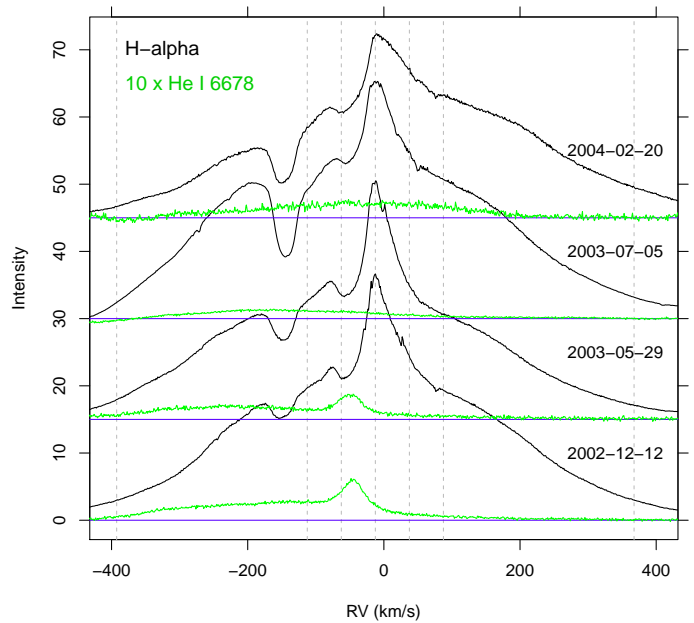}}
\caption{Selection of UVES spectra of $\eta$ Car, in the H$\alpha$ (black)
and He~I 6678 lines (green), over the years 2002-2004, as indicated
above each spectrum. Spectra
are continuum-subtracted, and shifted vertically for clarity.
The He~I line amplitude was enlarged 10 times to facilitate comparison
with H$\alpha$. Vertical dashed grey lines indicate the same velocities as in
Fig.~\ref{atlas-resid}.
\label{eta-car-spectra}}
\end{figure}

\begin{figure}
\resizebox{\hsize}{!}{
\includegraphics[bb=5 10 485 475]{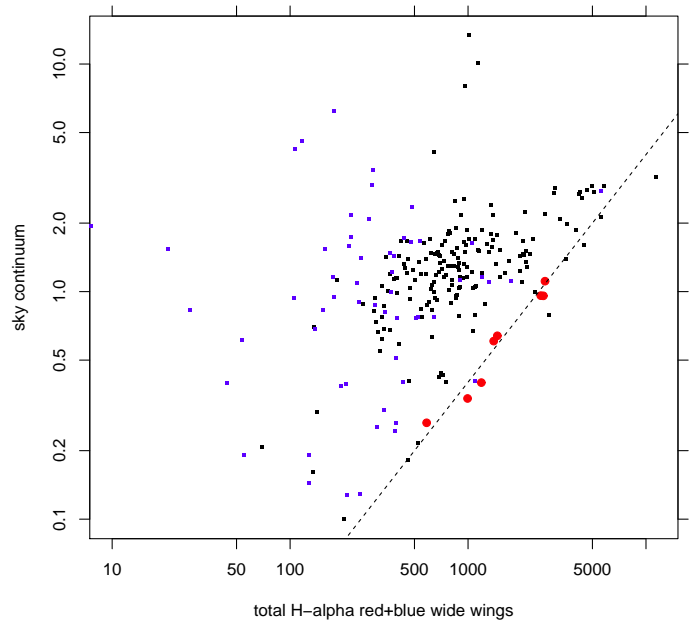}}
\caption{Sky continuum vs.\ total intensity in the wide H$\alpha$ wings.
Small dots refer to pure-sky fibres from setups HR15N (black) and HR14A
(blue), while big red dots are the (downscaled) continuum and wide wings
of $\eta$ Car. The dashed line is a constant factor of $4 \cdot 10^{-4}$.
\label{wings-continuum-etacar}}
\end{figure}
}

\begin{figure}
\resizebox{\hsize}{!}{
\includegraphics[bb=5 10 485 475]{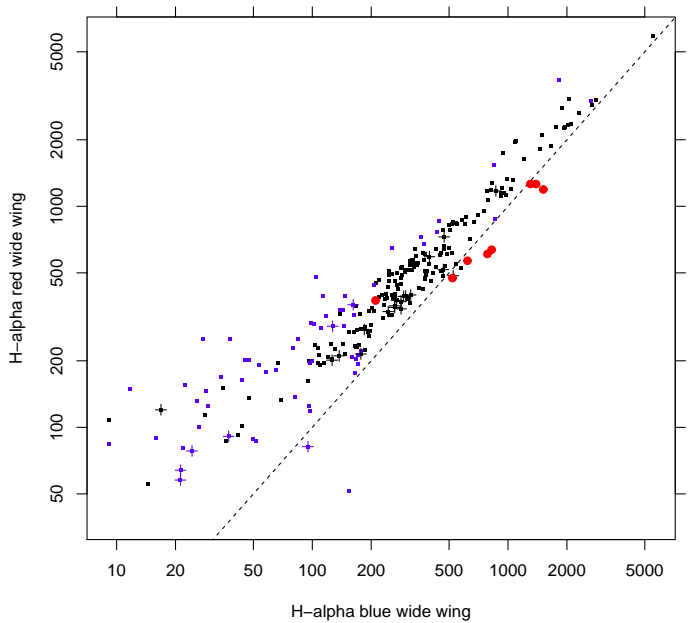}}
\caption{Intensity of red vs.\ blue wide H$\alpha$ wings. Datapoints from
pure-sky fibres of HR15N (black) and HR14A (blue) setups. The dashed
line is identity.
{The big red dots indicate the wide wings in the $\eta$~Car spectra.}
\label{wings-red-blue-wide}}
\end{figure}

{Figure~\ref{wings-red-blue-wide} shows that, as
already suggested from Figure~\ref{atlas-resid}, the intensities in the
blue and red parts of the wide H$\alpha$ wings are very well correlated with
one another.
Figure~\ref{wings-red-blue-wide} also shows that the intensity in the
blue wide wing is systematically lower than that in the corresponding red wing.
This is actually inconsistent with most of the measurements from
$\eta$~Car spectra (red datapoints); the signal-to-noise ratios of the
latter spectra is so high that the errors on the $\eta$~Car wing
measurements are negligible (comparable to the plotted symbol size), so
that the difference is highly significant. Also significantly different
are the velocities of the absorption reversal in the $\eta$~Car
spectra ($v \sim -150$ km/s) and in the nebular spectrum ($v \sim -125$ km/s)
of Fig.~\ref{atlas-resid}, panel $e$. Both discrepancies can be
reconciled, however, by assuming that the scattering dust is not at
rest. If the dust is expanding away from $\eta$~Car it will ``see'' its
H$\alpha$ emission as redshifted; then, depending on the dust relative motion
with respect to us, the reflected redshifted line may gain an additional
redshift, a blueshift, or no shift at all. For a spherically expanding
dust envelope, a total wavelength shift varying linearly with radial distance
is predicted. Therefore, Figure~\ref{wing-ratio-radius} shows the ratio
between the red and blue wide H$\alpha$ wings (increasing with redshift),
vs.\ radial distance from
$\eta$~Car: a clear correlation is seen for most datapoints, connected
smoothly with the actual $\eta$~Car values at zero distance.
No such correlation is instead found for distances $>5$ arcmin from
$\eta$~Car.
It should be
remarked that because of the variability found in the $\eta$~Car
H$\alpha$
profiles (red points), some scatter in the distribution of nebular values was
expected, in agreement with the handful of outliers seen in the
Figure. Therefore, the hypothesis of a reflection origin for the wide
H$\alpha$ wings is supported by our data only if the scattering dust is
radially expanding away from $\eta$~Car. On the other hand, we do not
find any reasons why the emission from genuine high-velocity gas should
show a pattern like that in Fig.~\ref{wing-ratio-radius}: therefore this
latter hypothesis should be discarded.
It should be also remarked that the reflecting dust is likely
to form a half shell expanding towards us, rather than away from us: in
the latter case, dust observed close to $\eta$~Car would
move away both from this star and from us, and the corresponding
redshift in the scattered emission would be larger than the redshift at
the shell edge further away, where dust moves tangentially with respect to us.
Therefore, a dust half-shell expanding away from us should show a negative
slope in the redshift-radius diagram, whereas the positive slope seen in the
diagram of Fig.~\ref{wing-ratio-radius} corresponds to a half-shell expanding
towards us.

\begin{figure}
\resizebox{\hsize}{!}{
\includegraphics[bb=5 10 485 475]{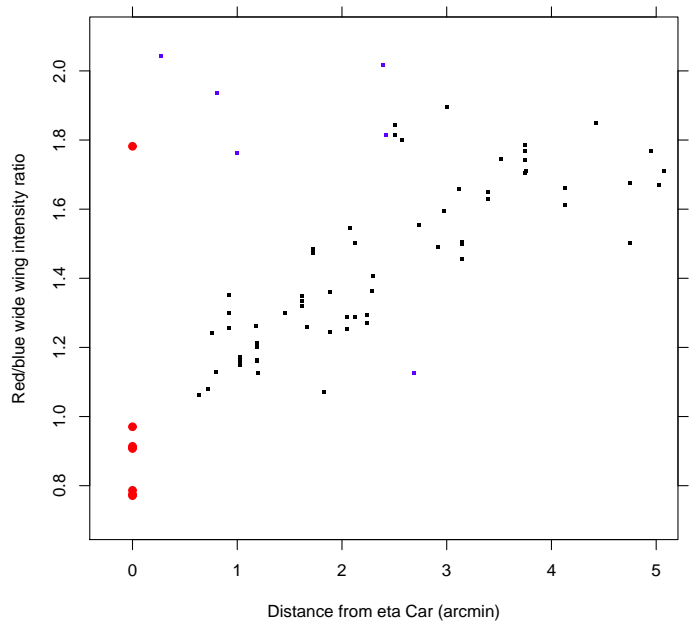}}
\caption{Ratio of red to blue wide H$\alpha$ wing intensities, in the
vicinity of $\eta$~Car, vs.\ radial distance.
Symbols as in Fig.~\ref{wings-red-blue-wide}.
\label{wing-ratio-radius}}
\end{figure}
}

\begin{figure}
\resizebox{\hsize}{!}{
\includegraphics[bb=5 10 485 475]{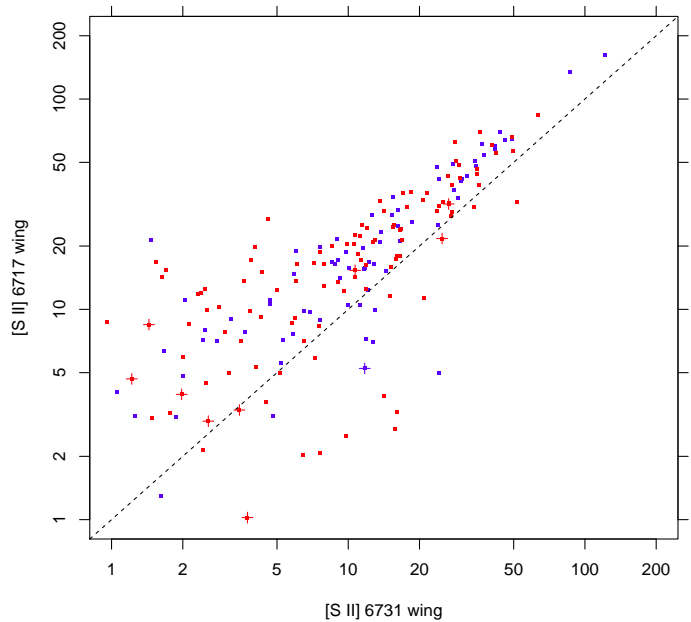}}
\caption{Intensity of high-velocity components in the [S II] 6717\AA\
line vs.\ those in the [S II] 6731\AA\ line.
The dashed line is identity. The average ratio, larger than one,
indicates low densities.
Symbols as in Fig.~\ref{ha-n2-rv}.
\label{wings-s2}}
\end{figure}

{All considered, the most reliable indicator of genuine diffuse
high-velocity gas ($v \sim \pm 100$ km/s),
found only at some places in the studied region,
is the occasional emission in the [S II] wings discussed above.
The doublet ratio
permits to derive the gas density of this high-velocity gas.
Figure~\ref{wings-s2} shows the
intensity in [S II] 6717\AA\ vs.\ [S II] 6731\AA: the former is on
average larger, indicating low densities ($\sim 100$ cm$^{-3}$).
}

\subsection{Spatial morphology}
\label{maps}

\begin{figure}
\resizebox{\hsize}{!}{
\includegraphics[bb=5 10 485 475]{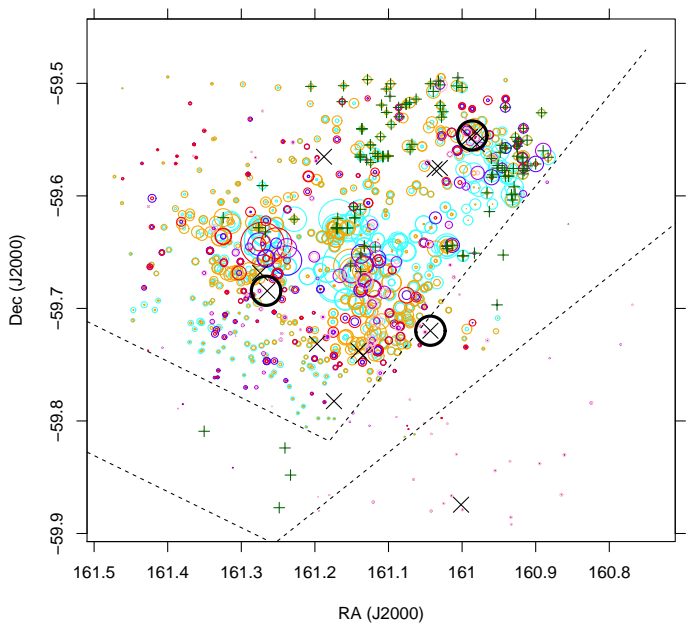}}
\caption{Spatial map of H$\alpha$ line intensity (fit normalization). The
radius of each circle is proportional to intensity. Blue (cyan, purple)
circles refer to the approaching component, while red (orange, pink)
circles refer to the receding component. Circle color codes as in
Fig.~\ref{ha-he-rv} and~\ref{ha-s2-rv}.
The oblique dashed lines outline the edges of the absorption lanes
visible in Fig.~\ref{dss}. Crosses indicate the positions of O/WR stars from
Walborn (1973).
Thick black circles indicate the positions of Trumpler~14
(upper), $\eta$~Car (middle left), and the Wolf-Rayet star WR25 (middle right).
The green '$+$' symbols indicate positions
of ``zero-velocity" gas, defined from Fig.~\ref{ha-rv}.
\label{map-ha-norm}}
\end{figure}

\begin{figure}
\resizebox{\hsize}{!}{
\includegraphics[bb=5 10 485 475]{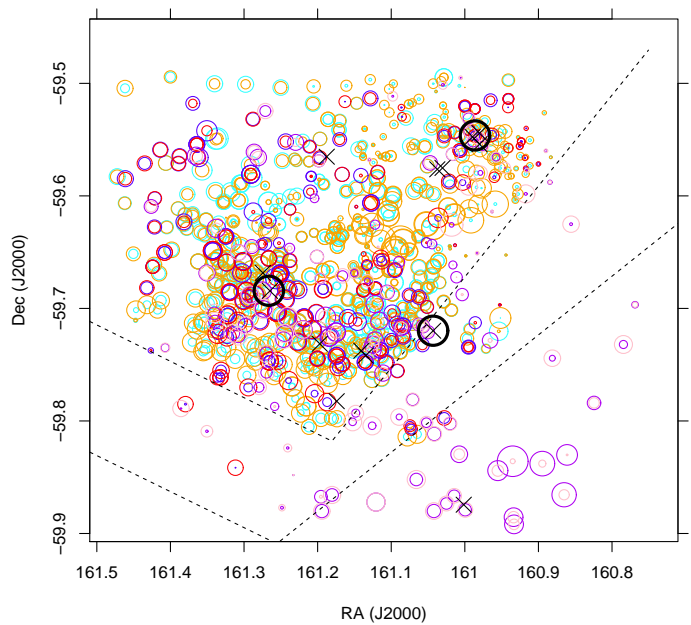}}
\caption{Map of RV of approaching and receding gas. Symbols as in
Fig.~\ref{map-ha-norm}. Circle size is proportional to $|RV-RV_{cm}|$,
with the largest circle corresponding to $|RV-RV_{cm}|=42.6$ km/s.
\label{map-ha-rv}}
\end{figure}

Important information on the emitting gas is derived from its spatial
distribution. Figure~\ref{map-ha-norm} shows a map of the intensity of
the two main components (blue/cyan/purple circles for the approaching
component, red/orange/pink circles for the receding one). The circle
radius is proportional to the best-fit normalization (scaled by exposure
time, and thus proportional to observed flux). The dark lane
position is indicated by the dashed lines. Crosses indicate O/WR
stars from Walborn (1973).
Thick black circles indicate the positions of Trumpler~14 center,
$\eta$ Car, and the Wolf-Rayet star WR25.
Finally, green plus symbols indicate positions of
``zero-velocity" gas identified in Figure~\ref{ha-rv}.

Figure~\ref{map-ha-norm} is very complex, but a few dominant features
can be observed. The distribution of surface brightness is not the same
for the approaching and receding components: this latter strongly dominates
especially in a region comprising $\eta$ Car and towards Southwest,
parallel to the southern dark lane. The approaching component dominates
North of this region, and also between it and the southern dark lane,
with impressive systematicity despite the low intensity of both
approaching and receding components there. Most of the zero-velocity gas
is found concentrated on two sides of Trumpler~14, again parallel to the
nearest dark lane. Overall, it is evident that plane-parallel
distributions prevail here over circular (spherical) configurations for
the bright material, as it was already clear for the dark, obscuring
material. Some sort of density stratification probably plays a very
important role in shaping the entire central region of the Carina Nebula.
We remark that the different intensity distributions for the two
components could have never been discovered using narrow-band H$\alpha$
images, since even narrow-band filters are not wavelength-selective
enough to image them one at a time.

Figure~\ref{map-ha-rv} is a map similar to Fig.~\ref{map-ha-norm},
but with circle radii
proportional to $|RV-RV_{cm}|$ for both approaching and receding
components. As already suggested from the discussion on Fig.~\ref{ha-rv}
there is no global velocity pattern; on smaller scales, we observe again
the near zero RVs on the two sides of Trumpler~14, and there are
indications that other patterns exist on similarly sized scales.
Therefore, the two main components here must be local to the inner
region of the Nebula, and not connected to the global expansion of the
whole Carina nebula, on scales of 20-30~pc, as already discussed above.

\begin{figure}
\resizebox{\hsize}{!}{
\includegraphics[bb=5 10 485 475]{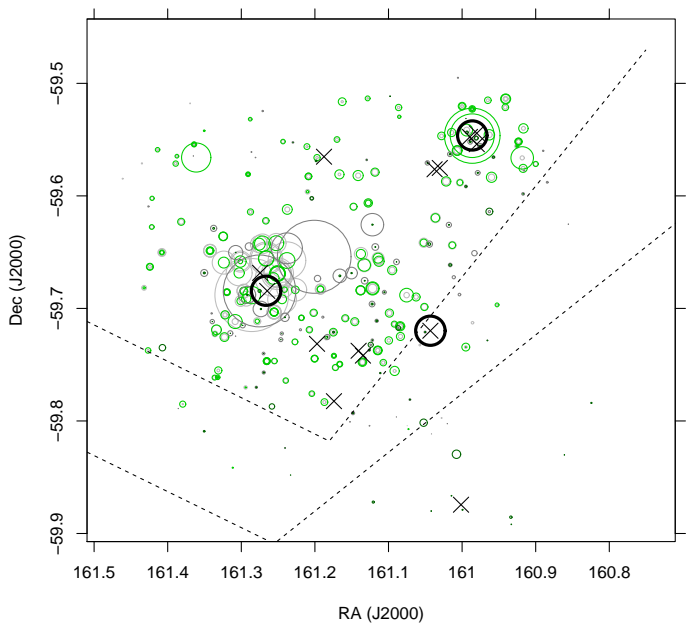}}
\caption{Map of the distribution of intensity in the wide H$\alpha$ wings
(grey) and sky continuum emission (green). Circle size is proportional
to intensity.  Darker grey circles indicate data from pure-sky
fibres using HR14A setup. Thick circles and crosses have the same
meaning as in Fig.~\ref{map-ha-norm}. The large grey circle closest to center
indicates {the very intense H$\alpha$ wings}
in the direction of the Keyhole nebula.
\label{map-continuum-wide}}
\end{figure}

Before discussing the local regions individually, we examine some more
distribution maps. {Figure~\ref{map-continuum-wide} illustrates the
distribution of wide H$\alpha$
wings intensity (grey circles, with radii proportional to intensity): it is
clear that they are found all around
$\eta$ Car, not only in the direction of the Keyhole Nebula.
}
The green circles instead indicate the
intensity of the sky continuum, which is distinctly measurable in some
of the pure-sky spectra, and reveals reflection nebulosity from dust in
the vicinity of bright stars. The fact that this continuum emission has
no features similar to the solar spectrum, and that its intensity varies
among sky spectra from the same OB, ensures that it is not originating
from scattered moonlight, but from dust reflection in Carina. The strongest
reflection nebulosity is evidently found in Trumpler~14, but also the
$\eta$ Car region shows clear detections of it.
{As discussed in the previous section},
we do not find a proportionality between the intensity
of the reflected sky continuum and that of the wide H$\alpha$ wings.

\begin{figure}
\resizebox{\hsize}{!}{
\includegraphics[bb=5 10 485 475]{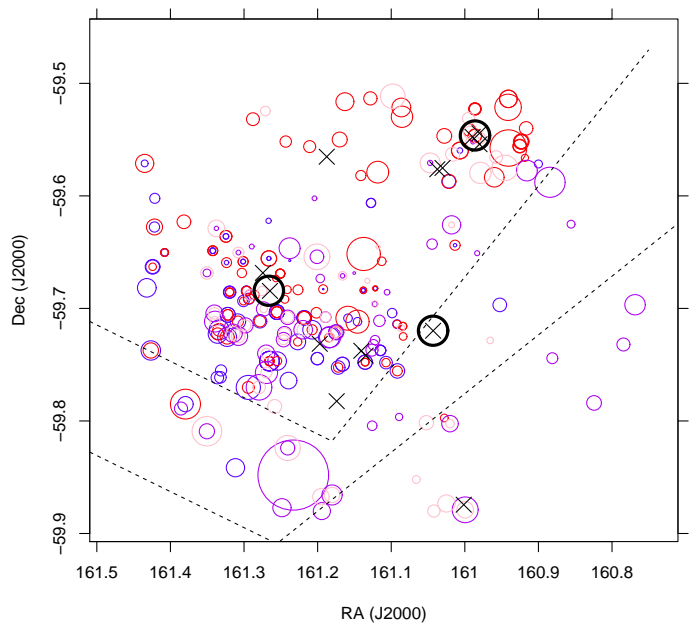}}
\caption{Map of the intensity ratio [N II] 6584\AA/H$\alpha$.
Symbols as in Fig.~\ref{map-ha-norm}, except for the (cyan/orange) data
from faint stars, not used here.
Circle size is proportional to ratio, with largest circle corresponding
to a value of 2.4.
\label{map-n2-ha}}
\end{figure}

Then, Figure~\ref{map-n2-ha} shows the spatial distribution of the
intensity ratio between the [N II] 6584\AA\ line and H$\alpha$.
However, since we found above that the
respective dynamics of hydrogen and nitrogen show that the two sets
of lines do not come from exactly the same gas but merely from adjacent
(or otherwise loosely connected) gaseous components, 
we consider their intensity ratio only in places where their RVs differ
by less than 3 km/s.
For constant ionization fractions of H and N, their intensity ratio is
an increasing function of gas temperature (Haffner et al.\ 2009).
Taking it at face value, under the hypothesis of constant H and N
ionization, the map would imply a wide range of temperatures for the
emitting regions (about a factor of 3, with exact values depending on
the assumed H and N ionization fractions). This is wider than the range
derived above from the line-width ratios (outliers excluded), and
moreover there is no correlation between the line-width ratios and the
N/H intensity ratios shown in Figure~\ref{map-n2-ha}. Therefore, it is
more likely that rather than a range in gas temperatures the ratios N/H
reflect here variations in the N (or H) ionization fraction. The Figure
shows then that this varies strongly, and in different ways for the
approaching and receding components (as found by Deharveng and Maucherat
1975): in the approaching component the N
ionization increases markedly south of $\eta$~Car (again, along a
region parallel to the southern dark lane), while in the receding
component it is more uniform, with some preference for higher values in
the northern parts. We will discuss an interpretation of this fact in
the next section.

\begin{figure}
\resizebox{\hsize}{!}{
\includegraphics[bb=5 10 485 475]{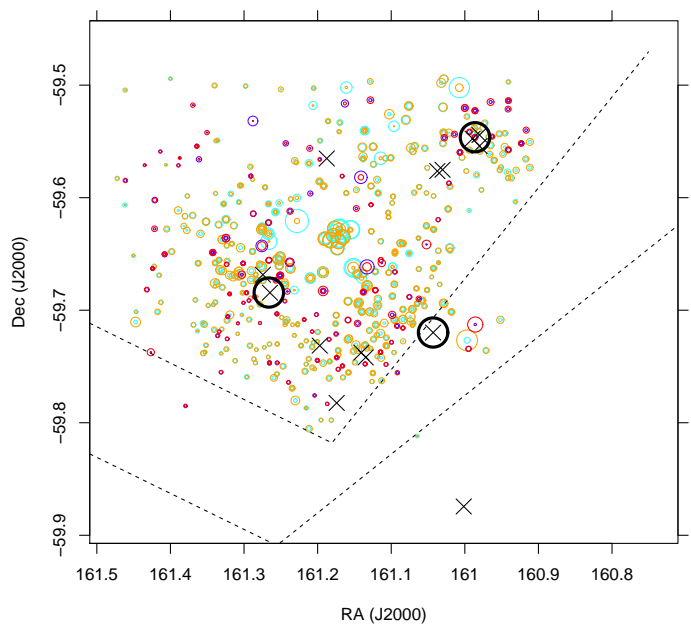}}
\caption{Map of the intensity ratio between [S II] 6731/6717\AA,
increasing with density.
Symbols as in Fig.~\ref{map-ha-norm}.
Largest circle corresponds to a ratio of 0.65.
Note the large values of this ratio near the Keyhole nebula, $\sim 3$
arcmin N-W of $\eta$ Car.
\label{map-s2-ratio}}
\end{figure}

\begin{figure}
\resizebox{\hsize}{!}{
\includegraphics[bb=5 10 485 475]{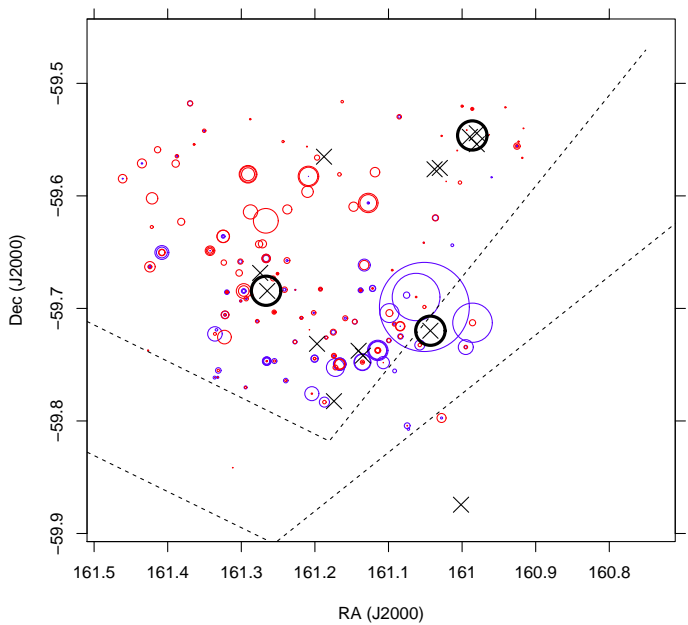}}
\caption{Map of intensities of emission in the far wings of [S II] 6717.
Symbols as in Fig.~\ref{map-ha-norm}.
\label{map-s2-wings}}
\end{figure}

Figure~\ref{map-s2-ratio} shows the distribution of the intensity ratio
between [S II] 6731/6717\AA\ (increasing with density), of which the
largest values are especially found in the region surrounding the Keyhole
nebula.
Finally, Figure~\ref{map-s2-wings} shows the intensity distribution of
the high-velocity [S II] emission occasionally found (see
Fig.~\ref{atlas-resid}), which is especially frequent as a receding
component North of $\eta$ Car, and as an approaching component next to
the Western lane, and may trace high-speed shocks (Hartigan et al.\ 1999).
A closer examination of shock diagnostics will be presented in
Section~\ref{lineratios} below.

\subsection{Position-velocity diagrams of shell-like structures}
\label{shell}

\begin{figure}
\resizebox{\hsize}{!}{
\includegraphics[bb=5 10 485 475]{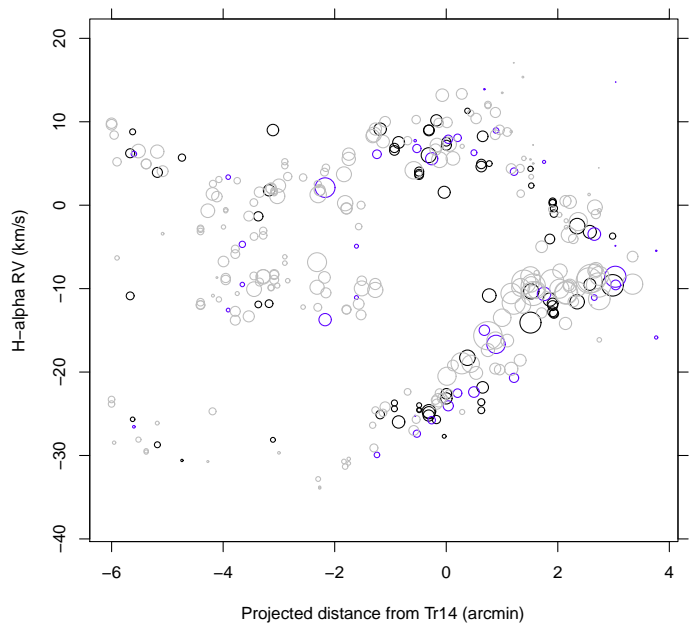}}
\caption{Position-velocity diagram for H$\alpha$ in the neighborhood of Trumpler~14.
Circle size is proportional to line intensity. Color codes as in
Fig.~\ref{ha-rv}.
The distance origin is taken at the Trumpler~14 position indicated with
the big black circle in the upper right of Fig.~\ref{map-ha-norm}.
Positive distances are towards S-W; the dark lane lies at position $\sim
+3$~arcmin. Direction of the observer is towards the bottom, at negative RVs.
\label{dist-rv-tr14}}
\end{figure}

As discussed above with reference to Figs.~\ref{map-ha-norm}
and~\ref{map-ha-rv}, velocity
structures seem to exist at local scales, rather than on
{the scale of the entire studied region.}
It was also remarked that plane-parallel geometries are dominant
over circular ones, so it becomes natural to look for velocity
structures along preferred directions, rather than along the radial
direction inside circular regions. The first example shown here in
Figure~\ref{dist-rv-tr14} is a position-velocity diagram in the region
surrounding the Northern cluster Trumpler~14, already mentioned in
connection to the zero-velocity gas.
In Fig.~\ref{dist-rv-tr14}, the abscissa reports position along an axis
at position angle of $110^{\circ}$ (from North, clockwise), passing
though Tr~14 center. The chosen position angle reflects the spatial alignment
of zero-velocity datapoints.
The circle radius is proportional to H$\alpha$ intensity
(fit normalization), and approaching and receding components are
both indicated with the same colors. The pattern indicates some sort of
radial expansion, approximately centered in Tr~14. The geometry is not
spherical, however, but distorted, probably because the medium into
which expansion is taking place has strongly non-uniform density.
Some non-spherical expansion geometry was already suggested by
Deharveng and Maucherat (1975).
We find an abrupt discontinuity in the RV pattern for the approaching
component at projected distances between (-2,-1) arcmin: to the left of
this position, the approaching gas RVs flatten out to a nearly constant
value close to -10 km/s, as if the expanding gas had hit a much denser
layer of material traveling at that speed, and was basically stopped and
forced to co-move with it.

\begin{figure}
\resizebox{\hsize}{!}{
\includegraphics[bb=5 10 485 475]{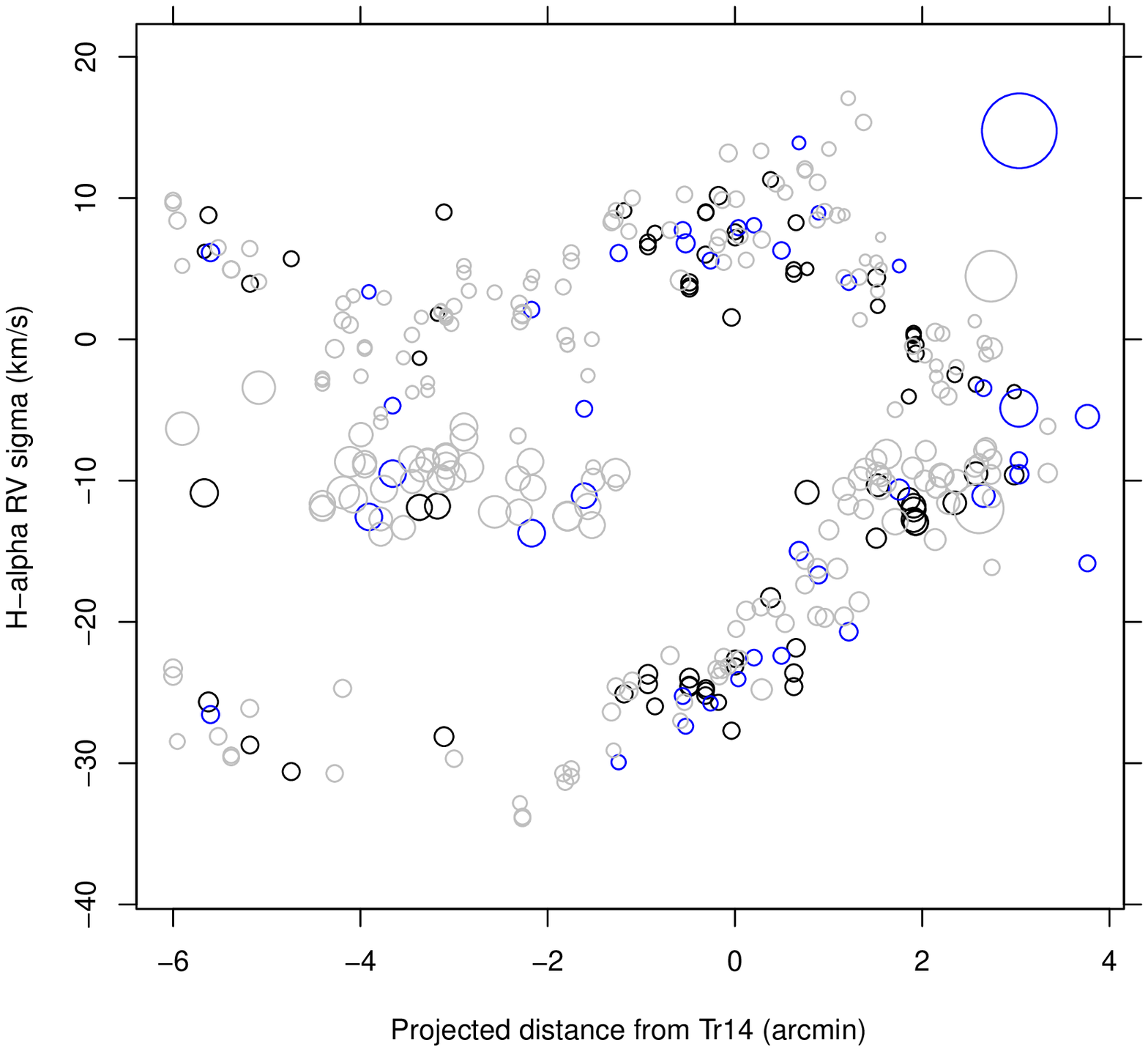}}
\caption{Same as Fig.\ref{dist-rv-tr14}, except that circle size is
here proportional to line width $\sigma$.
\label{dist-rv-sigma-tr14}}
\end{figure}

\begin{figure}
\resizebox{\hsize}{!}{
\includegraphics[bb=5 10 485 475]{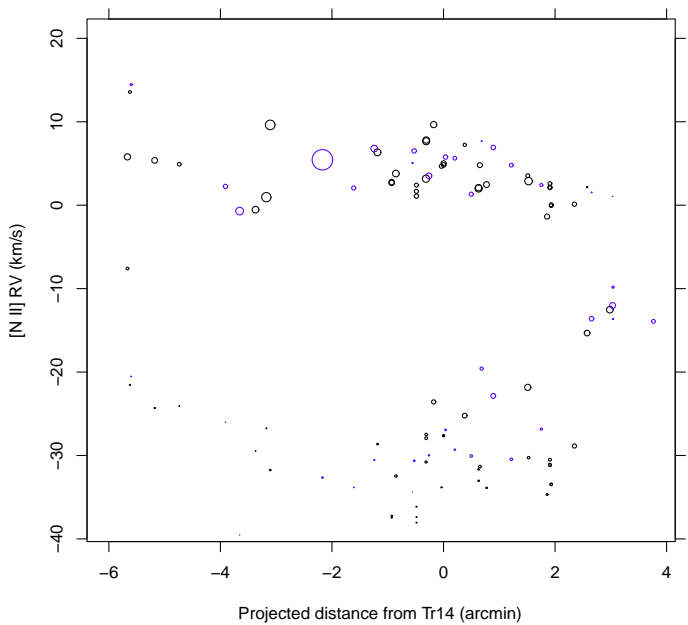}}
\caption{Same as Fig.\ref{dist-rv-tr14}, but using the [N II] 6584\AA\ line.
\label{dist-rv-n2-tr14}}
\end{figure}

Additional evidence for such an interaction having occurred between the
expanding gas and another layer may come from the position-velocity
diagram of Figure~\ref{dist-rv-sigma-tr14}, which differs from the
previous Figure in that the circle size is proportional to H$\alpha$ line width
$\sigma$, and thus to turbulent speed (or temperature): it is clear from
this latter diagram that the line width increase systematically and
significantly in the ``stopped" gas, likely as a result of a collision
with a massive obstacle. Also interesting is the corresponding
position-velocity diagram using [N II] instead of H$\alpha$: in fact, in this
region are found many of the outliers seen in Fig.~\ref{ha-n2-rv}, where
the H$\alpha$ RV was near $RV_{cm}$ while the RV of [N II] emission was
similar to the rest of the approaching gas. Analogously,
Figure~\ref{dist-rv-n2-tr14} shows that the expansion of the Tr~14 shell
seen in the [N II] line is very different than seen in H$\alpha$: the shell
morphology is barely discernable, and more importantly the [N II]
emission does not indicate any slowed-down expansion where H$\alpha$ does,
but continues to show large approaching velocities. Obviously, the
[N II] emitting gas is different, and in this particular case lying beneath the
gas emitting H$\alpha$, and has yet to reach the space region where
slowing-down occurs.

\begin{figure}
\resizebox{\hsize}{!}{
\includegraphics[bb=5 10 485 475]{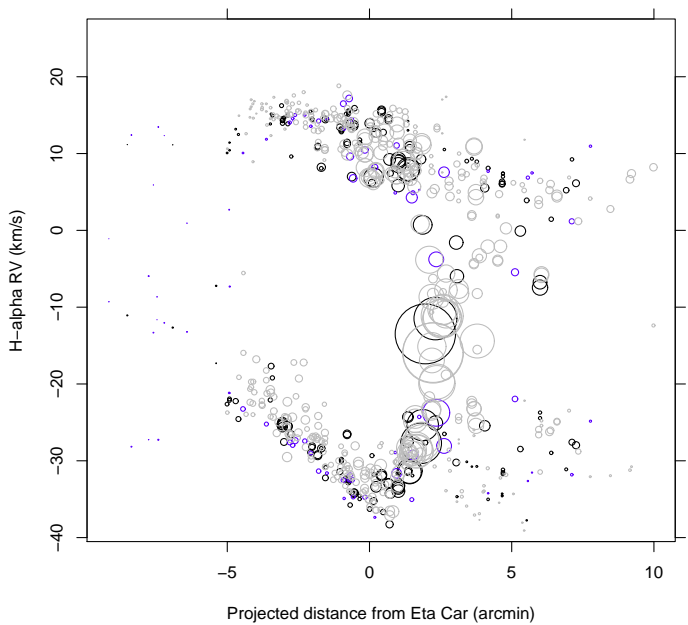}}
\caption{Same as Fig.\ref{dist-rv-tr14}, but centered on $\eta$ Car.
Positive distances are towards N-W direction from $\eta$ Car.
The Keyhole nebula lies at position $\sim +3$~arcmin.
\label{dist-rv-etacar}}
\end{figure}

\begin{figure}
\resizebox{\hsize}{!}{
\includegraphics[bb=5 10 485 475]{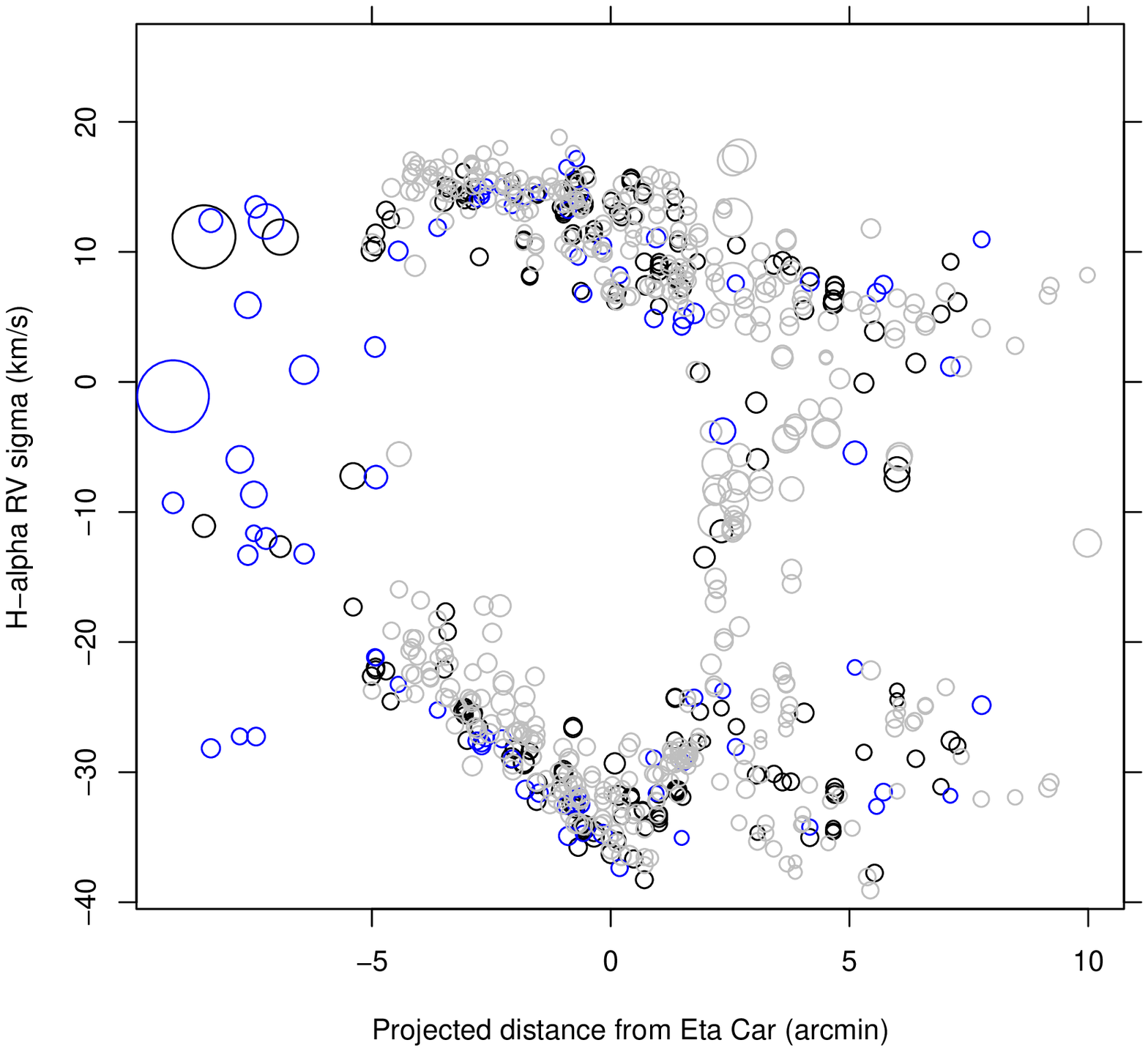}}
\caption{Same as Fig.\ref{dist-rv-etacar}, except that circle size is
here proportional to line width $\sigma$.
\label{dist-rv-sigma-etacar}}
\end{figure}

Also worth investigating is the region surrounding $\eta$ Car, as
remarked in the discussions above on spatial distributions.
Figure~\ref{dist-rv-etacar} is a velocity-position diagram like
Fig.~\ref{dist-rv-tr14} but centered on $\eta$ Car, towards position
angle $35^{\circ}$. Here too a sort of distorted, incomplete shell is
suggested by the diagram: the incompleteness towards South is due to the
strong obscuration by the southern dust lane (left of distance of -5 arcmin
in the Figure). At projected distance $\sim 2.5$ arcmin the shell
border, converging towards $RV_{cm}$, is clearly visible; the same
datapoints show increased H$\alpha$ line intensities (larger circles in the
Figure), as expected because of projection effects.
This position nearly coincides with the Keyhole nebula.
To the right of this
position, H$\alpha$ emission is much weaker, but the RV pattern suggests the
existence of another shell towards NNW,
possibly centered on the O3 star HD93250.
Near position $\sim 2.5$ arcmin, an interaction between the
two shells might be taking place, and contributing to the peculiarities of
the Keyhole nebula summarized in Smith and Brooks (2008).
Such an interaction might be diagnosed by an increased gas turbulence,
which as can be understood from Fig.~\ref{ha-he-sigma} is an important
contributor to measured line widths.
To investigate this possibility we
show in Figure~\ref{dist-rv-sigma-etacar} the same diagram with circle
sizes proportional to H$\alpha$ line widths: these are generally small
(symbols adopt the same scaling factor
here as in Fig.~\ref{dist-rv-sigma-tr14}), but increase precisely in the
group of datapoints around position $\sim 2.5$ arcmin and RV$\sim -10$
km/s, i.e.\ the putative interaction region between the shells. This
might be an indication of ongoing shell collision and merging.

\begin{figure}
\resizebox{\hsize}{!}{
\includegraphics[bb=5 10 485 475]{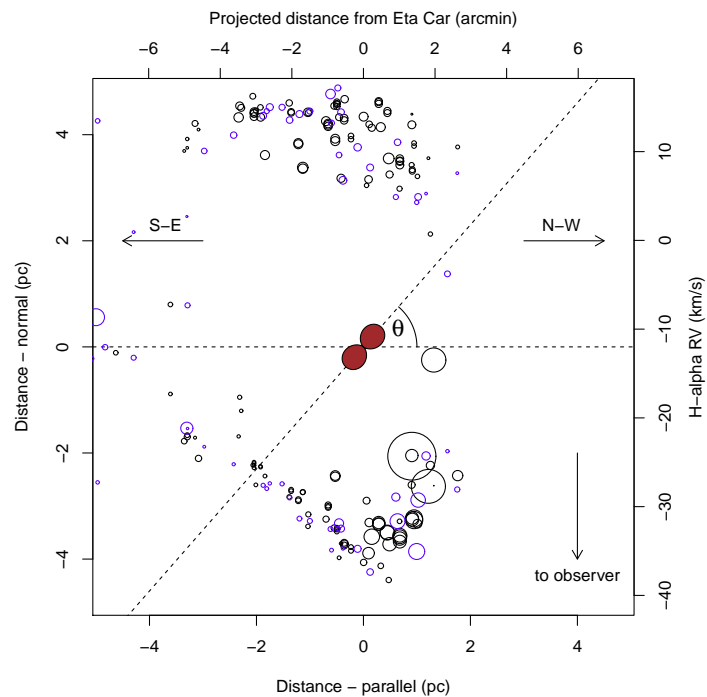}}
\caption{Reconstructed geometry of the $\eta$~Car shell.
Bottom (left) axis indicates distances parallel (normal) to the plane of sky.
Top and right
axes indicate the original measurements, as in Fig.~\ref{dist-rv-etacar}.
Symbols as in this latter Figure, except that here circle size is
proportional to the H$\alpha$/[N II] intensity ratio.
The brown double-lobe structure
near the center represents the position and actual inclination of the
Homunculus nebula (size is not to scale).
The horizontal dashed line, which defines angle $\theta=0$, is parallel
to sky plane, the oblique line indicates the Homunculus polar axis.
\label{dist-n2ha-etacar}}
\end{figure}

\begin{figure}
\resizebox{\hsize}{!}{
\includegraphics[bb=5 10 485 475]{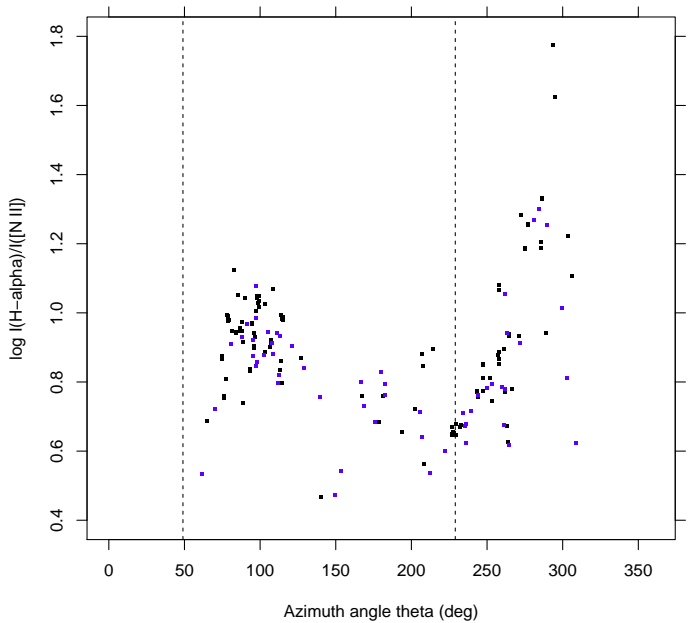}}
\caption{Intensity ratio H$\alpha$/[N II] vs.\ azimuthal angle $\theta$,
defined in Fig.~\ref{dist-n2ha-etacar}. Symbols as in Fig.~\ref{ha-norm}.
Vertical dashed lines indicate the projections of the Homunculus polar
axis.
\label{azimut-n2ha-etacar}}
\end{figure}

The $\eta$~Car shell presents a rich phenomenology which deserves a
closer examination.
As evident from Fig.~\ref{map-n2-ha}, in the
half-plane S-E of $\eta$~Car the [N II]/H$\alpha$ intensity ratio is systematically
larger in the blueshifted component than in the redshifted one.
Moreover, under the rough zero-order
assumption of spherical expansion, we can transform the line RV into a
distance from the expansion center (i.e., $\eta$~Car), with a best-guess
conversion factor such that the shell is as round as possible in shape.
The result of this transformation is shown in
Figure~\ref{dist-n2ha-etacar}, where axes are scaled to parsecs
using the known Nebula distance; the original measurements
(sky-projected distance in arcmin and RV) are shown at top and right
borders, respectively.
Only components which differ in RV by less than 3~km/s are used to
compute the [N II]/H$\alpha$ ratio (proportional to symbol size in
Fig.~\ref{dist-n2ha-etacar}).
We know the inclination angle to the
line of sight of the Homunculus Nebula surrounding $\eta$~Car
(whose two lobes are indicated in Fig.~\ref{dist-n2ha-etacar} in brown color)
from the very detailed studies of Davidson et al.\ (2001) and Smith (2006).
It turns out
that the H$\alpha$/[N II] ratio varies in a systematic way with azimuthal
angle $\theta$, defined in the Figure (and increasing counterclockwise),
with $\theta=49^{\circ}$
corresponding to the projection of the Homunculus polar axis in the N-W
direction (with a component away from us),
and $\theta=229^{\circ}$ corresponding to its
projection towards S-E (with a component towards us).
As a function of $\theta$, the H$\alpha$/[N II] ratio is shown in
Figure~\ref{azimut-n2ha-etacar}: fairly well-defined minima are found
corresponding to each of the poles.
In the next section we will discuss that the H$\alpha$/[N II] intensity ratio
may be considered a good indicator for the ionization parameter $q$.
Thus, the minima in Fig.~\ref{azimut-n2ha-etacar} also correspond to minima in
the ionization parameter $q$:
gas lying along the Homunculus polar axes is
therefore irradiated with much less UV flux than gas in the equatorial
directions ($\sim 30$ times less). This is possible if the total column
density in the matter surrounding $\eta$~Car is not isotropic, but much
higher near the polar axes.  This picture agrees very well with
the findings of Smith (2006), that most of the matter forming the
Homunculus was originally ejected by $\eta$~Car at high latitudes, and
did not get its shape by the effect of a circumstellar torus.
If this latter picture was the case, we would have observed a brighter
UV irradiation near the poles, i.e.\ the opposite of what our data
suggest. Our picture also implies that $\eta$~Car is by far the dominant
ionizing source inside its own shell, and probably throughout the entire
Trumpler~16 cluster.

Such directionality in the ionization pattern around $\eta$~Car, and the
shape itself of the Homunculus, would
suggest a polar-angle dependence also for the mechanical energy output
of $\eta$~Car winds. If this is true, the fact mentioned above that the
(turbulent) line widths increase near the inter-shell boundary, but not along
the $\eta$~Car polar axis, might then suggest that winds from massive stars
are a secondary contributor to turbulence in H$_{\rm II}$ regions, in agreement
with Krumholz and Burkhart (2015).

\begin{figure}
\resizebox{\hsize}{!}{
\includegraphics[bb=5 10 485 475]{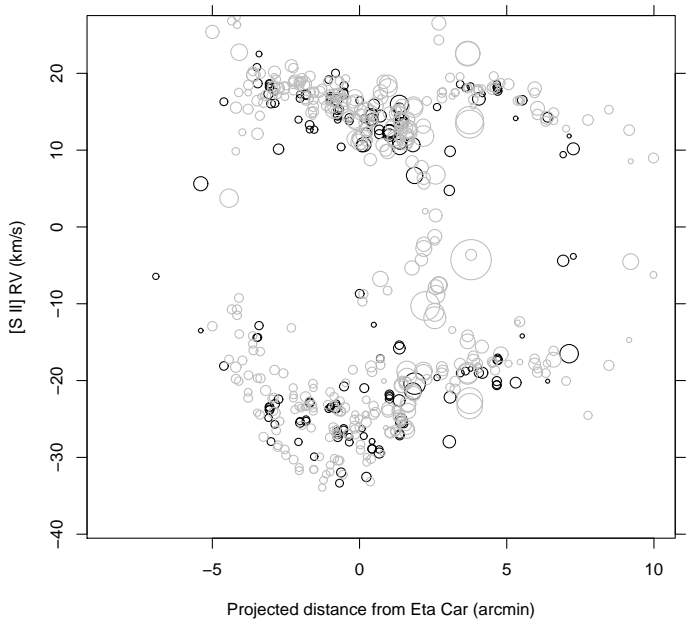}}
\caption{Same as Fig.\ref{dist-rv-etacar}, except that circle size is
proportional to [S II] 6731/6717 intensity ratio.
\label{dist-rv-s2ratio-etacar}}
\end{figure}

We have also examined if a similar pattern exists for the gas density,
as measured from the [S II] 6731/6717 line ratio. The corresponding diagram
is shown in Figure~\ref{dist-rv-s2ratio-etacar}, with circle sizes
proportional to the [S II] 6731/6717 intensity ratio (increasing with
density). Here, no dependence on closeness to Homunculus polar axis is
found; instead, the line ratio and gas density tends to increase
systematically from S-E towards N-W (left to right in the Figure) for
projected distances $<3-4$ arcmin, with
the highest densities found near the N-W shell edge, corresponding to
the Keyhole nebula, as remarked above.

\begin{figure}
\resizebox{\hsize}{!}{
\includegraphics[bb=5 10 485 475]{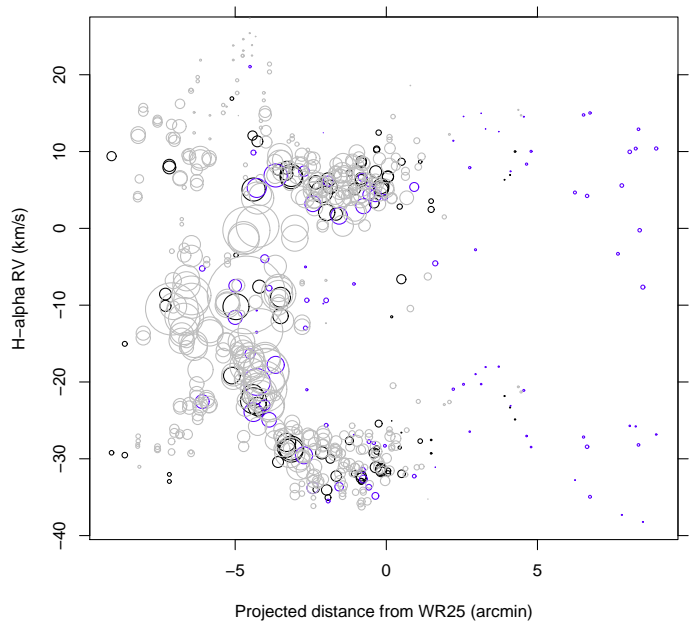}}
\caption{Same as Fig.\ref{dist-rv-tr14}, but centered on WR25.
\label{dist-rv-sw}}
\end{figure}

\begin{figure}
\resizebox{\hsize}{!}{
\includegraphics[bb=5 10 485 475]{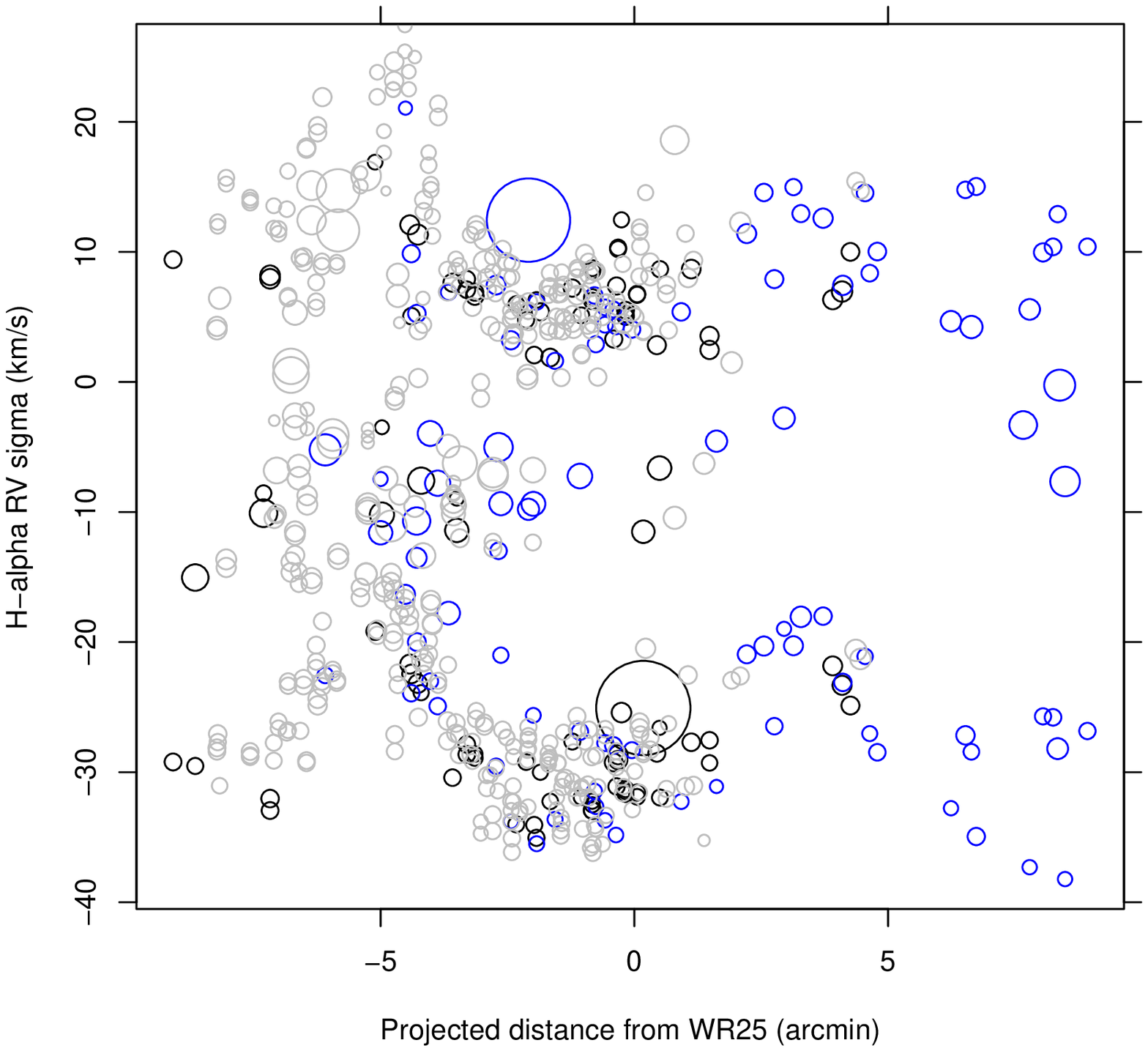}}
\caption{Same as Fig.\ref{dist-rv-sw}, except that circle size is
here proportional to line width $\sigma$.
\label{dist-rv-sigma-sw}}
\end{figure}

\begin{figure}
\resizebox{\hsize}{!}{
\includegraphics[bb=5 10 485 475]{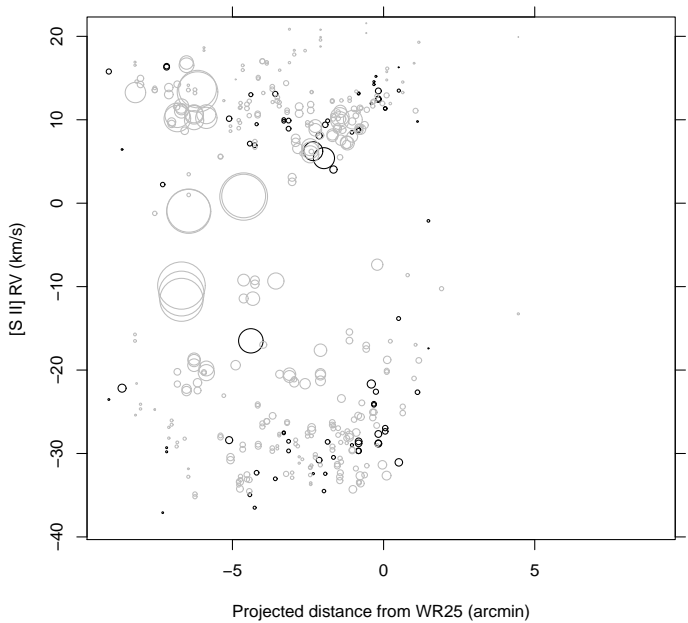}}
\caption{Same as Fig.\ref{dist-rv-sw}, but now using [S II] 6731\AA\
intensity.
\label{dist-rv-s2-sw}}
\end{figure}

Finally, Figure~\ref{dist-rv-sw} suggests the existence of a third
prominent shell, centered on the Wolf-Rayet star WR25 ("WR25 shell"),
with position angle $150^{\circ}$.
Here too, the southern part of the shell
is hidden behind the (western) dark dust lane, while at position $\sim -2.5$
arcmin the shell boundary is seen in the diagram, with enhanced
H$\alpha$ luminosity as above. Also in this case,
to the left of that boundary traces of another shell can be seen (presumably
the same shell North of $\eta$~Car as mentioned above). Here again we
test using line widths if interactions may be present:
Figure~\ref{dist-rv-sigma-sw} suggests that this may indeed be the case,
with increased line widths at several places near $RV_{cm}$. Also the
much increased intensity of the [S II] lines at some places in this
region, shown in Figure~\ref{dist-rv-s2-sw},
might be indicative of localized shocks at the interface between the two
shells (Hartigan et al.\ 1999).

All shells show a distorted shape, indicative of density gradients of the
material into which they expand. For this reason it is difficult to define
a typical ``radius" for each of them.
The maximum RV interval $\Delta RV$ at shell center
is instead clearly defined from the position-velocity diagrams, and found
to be $\Delta RV \sim 35$ km/s for the Tr~14 and WR25 shells, and
a larger $\Delta RV \sim 45$ km/s for the $\eta$~Car shell.
The approximate sizes of the shells ($\sim 5$ arcmin radius) and the observed
expansion speed allow to estimate a timescale, which turns out to be
very short, $\sim 1.5 \cdot 10^5$ years for the $\eta$~Car shell. Since this
is much shorter than the estimated ages of the massive stars in Carina,
it is likely that a much larger amount of material should be present
exterior to the bright nebulosity. Detection of a complex velocity
structure in the interstellar absorption Na I D lines
was indeed already known towards the center of Carina (Walborn et al.\ 2007 and
references therein), and
demonstrates the existence of substantial amounts of neutral gas in addition
to the predominantly ionized gas studied here. The Gaia-ESO data on the Na I
interstellar lines in Carina will be studied in a future work.

\subsection{Line-ratio diagrams}
\label{lineratios}

\begin{figure}
\resizebox{\hsize}{!}{
\includegraphics[bb=5 10 485 475]{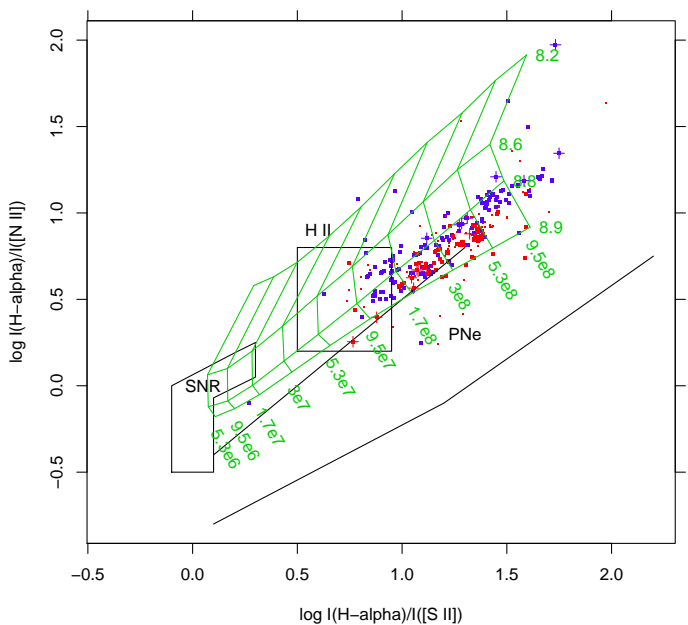}}
\caption{Plot of intensity ratios H$\alpha$/[N II] vs.\ H$\alpha$/[S II]. Symbols as
in Fig.~\ref{ha-he-rv}. Tiny dots indicate measurements where the
absolute RV difference between H$\alpha$ and [N II], or H$\alpha$ and [S II], is
larger than 3~km/s.
Black lines/polygons indicate typical loci of
Supernova Remnants (SNR), Planetary Nebulae (PNe), and Galactic H$_{\rm
II}$ regions
(H II), as labeled.
The green line grid is part of the {\sc CLOUDY} model grid shown in
Viironen et al.\ (2007), over a range of O/H abundances and ionization
parameters $q$, as labeled.
\label{ha-n2-ha-s2}}
\end{figure}

\begin{figure}
\resizebox{\hsize}{!}{
\includegraphics[bb=5 10 485 475]{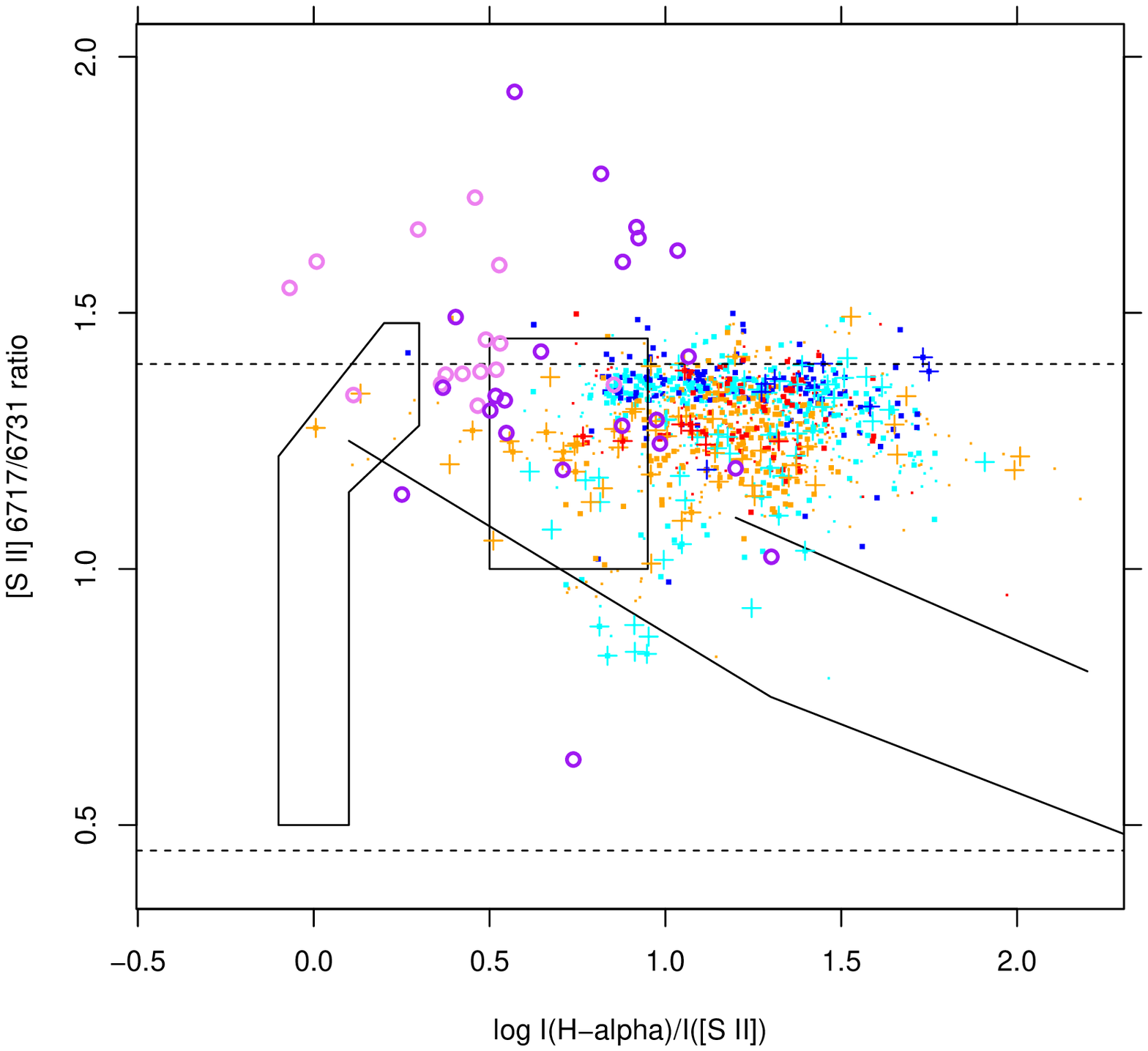}}
\caption{[S II] $\lambda\lambda$ 6717/6731 intensity ratio vs.\
H$\alpha$/[S II]
ratio. Circles refer to intensity ratios in the narrow wings of [S II]
and H$\alpha$, when present (violet: approaching; purple: receding component).
Other symbols as in Fig.~\ref{ha-s2-rv}. Solid lines/polygons indicate
SNR, H$_{\rm II}$ and PNe loci as in Fig.~\ref{ha-n2-ha-s2}. Dashed horizontal
lines enclose the range where the [S II] 6717/6731 ratio is an useful
measure of density (increasing from top to bottom).
\label{s2-ratio-ha-s2}}
\end{figure}

In this section we discuss briefly how line ratios in our dataset
compare with measurements from other nebular regions, including H$_{\rm
II}$
regions, Planetary Nebulae (PNe), and Supernova Remnants (SNR). Line-ratio
diagrams involving the lines used in this work were presented by
Sabbadin et al.
(1976), and then by Riesgo and L\'opez (2006) and Viironen et al.\ (2007).
They showed that these different object classes populate different
regions in the diagrams, with minimal mutual overlap. We show in
Figure~\ref{ha-n2-ha-s2} the intensity ratios of H$\alpha$/[N II] vs.\
H$\alpha$/[S II].
As is clear from Figs.~\ref{ha-n2-rv} and~\ref{ha-s2-rv}, however, some
sightlines have discrepant RVs between H$\alpha$ and [N II] ([S II]):
whenever the RV discrepancy was more than 3~km/s in absolute value the
line-ratio diagram, mixing intensities of emissions from slightly
different nebular layers, must be treated with caution: corresponding
datapoints are therefore shown with tiny dots.
The black lines and polygons shown in Fig.~\ref{ha-n2-ha-s2} indicate
typical loci of H$_{\rm II}$ regions, PNe and SNR. None of our datapoints falls
in the SNR region, where heating is dominated by shocks. The bulk of
datapoints lies just above the PNe wide strip (note however that the
limits of these loci are only indicative: see the much wider
distributions of PNe points in Riesgo and L\'opez 2006). Nearly half of
our datapoints fall outside the ``nominal" H$_{\rm II}$ box in the Figure: this
same pattern was already found for extragalactic H$_{\rm II}$ regions by
Viironen et al.\ (2007). These latter authors compare their datapoints with
grids of nebular ionization models, and show that such a pattern is
indeed expected for a fixed [O/H] abundance ratio, but varying
ionization parameter. In particular, our datapoints agree with oxygen
abundance of $8.7 < 12+\log(O/H) < 8.9$, and ionization parameter in the
range $6 \cdot 10^7 < q < 10^9$. Both H$\alpha$/[N II] and H$\alpha$/[S II] increase
with $q$ in the given range. The fact that most of our
datapoints fall at higher H$\alpha$/[S II] values than typical Galactic
H$_{\rm II}$
region is likely caused by the ionization parameter $q$ in Carina
exceeding typical H$_{\rm II}$ values elsewhere in the Galaxy.

Another useful diagram introduced by Sabbadin et al.\ (1976) is the
[S II] $\lambda\lambda$ 6717/6731 intensity ratio vs.\ H$\alpha$/[S II]
(Figure~\ref{s2-ratio-ha-s2}). Here again loci of H$_{\rm II}$ regions, PNe and
SNR are indicated. This diagram contains more datapoints than that of
Fig.~\ref{ha-n2-ha-s2} since it is based on three lines only, not
involving [N II]. In addition, we include measurements (circles) from narrow
line wings (section~\ref{resid}), which are sometimes very evident in [S II].
Plus signs indicate instead zero-velocity gas, which may lie at the
interface between adjacent shells. Fig.~\ref{s2-ratio-ha-s2} shows that
a handful of datapoints from these two latter categories lie
in the SNR region or very close to it, suggesting that shocks may indeed
be present, but only in very few places in the nebula, including the
shell boundaries.

\section{Discussion}
\label{discuss}

The new observations of nebular emission lines in the central regions of
the Carina Nebula have put into evidence a rich variety of new facts.
The emitting gas is highly differentiated in its kinematic properties:
the hydrogen-emitting and forbidden-line-emitting gas have in general
similar motion, but may also sometimes show highly contrasting
velocities, as in the case of the Tr~14 shell. {The low-level
high-velocity emission was studied here with better detail than in
previous studies, thanks to the very large S/N of the new data.

We have presented new evidence showing that the wide H$\alpha$ wings arise
from reflection of the emission of $\eta$~Car off dust, strengthening
results from previous works. However, to be entirely consistent with our
observations, such dust cannot be stationary, but expanding away from
$\eta$~Car. It should be remarked that $\eta$~Car is spectrally variable
on timescales shorter than the light-crossing time ($\sim 11$~yr) of
the putative dust shell radius of $\sim 5$ arcmin. Regarding longer-term
variability, we are fortunate that such a 11-yr delay between the direct
and reflected $\eta$~Car emission coincides nearly with the time
difference between our observational
data sets (2002-2004 for $\eta$~Car UVES spectra; Apr.\ 2014 for the
bulk of our HR15N nebular spectra): this makes that the comparison
between $\eta$~Car and its reflected emission, e.g.\ in
Fig.~\ref{wing-ratio-radius}, is unaffected by long-term variations.

The narrow excess emission seen in the H$\alpha$ fit residuals between
absolute RVs of 50-100~km/s is of uncertain significance. More
significant is instead the occasional excess emission in the [S II]
lines, in the same RV range.
Such enhanced high-velocity [S II] emission appears at almost random
places; it might be originated where the fast OB star winds
encounter local overdensities in the ambient medium, with formation of
shocks, as suggested by line-ratio diagrams.
The reason why this phenomenon happens preferentially around $|RV|
\sim 100$ km/s is also unclear;
}
it might be only a coincidence that this is close to the
sound speed in the million-degree hot plasma which is found diffusely in
the same region (Townsley et al.\ 2011), and which is supposed to be
also heated by shocks caused by the OB stars winds. A detailed
theoretical investigation of the complex observed phenomena in this complex
region would be certainly desirable.

{
The two main expanding layers at $RV \sim RV_{cm} \pm 20$ km/s
were known long before this work, as mentioned in the introduction.
The $\sim 20^{\prime} \times 20^{\prime}$ spatial region studied here, in the
brightest, central part of Carina, is nearly the same as that studied by
Deharveng and Maucherat (1975) using lesser-quality data,
and covers about one-third of that
studied by Meaburn, Lopez and Keir (1984), who however analyzed
Fabry-Perot spectra only along two lines crossing Carina.
Our findings from the previous section support a local origin for
the bulk of approaching/receding line emission, from few
discrete expanding shells of gas surrounding the most massive stars.
In order to place these findings in the appropriate context, we may
refer to the larger-scale images in Smith and Brooks (2007), and their
Figure~5$b$ in particular, showing an H$\alpha$/[O III] flux-ratio image.
This figure suggests strongly that the entire Carina nebula is composed by a
collection of shells and cavities, of projected sizes in the range
15-60~arcmin, and thus much larger than the shells found in this work.
Both Trumpler~14 and~16 are found in the northern lobe of a bipolar
cavity (see also Smith et al.\ 2000), apparently split in two by the dark
dust lanes. The same image does not show sub-shells of smaller scale,
corresponding to those we have described in the previous section (all
falling in the same northern lobe of the bipolar cavity). To reconcile
our results with the larger-scale gas bubbles, we hypothesize that our
shells have already started to merge
together, producing the larger structures of the bipolar cavity.
This structure, with each of its lobes having an asymmetrical shape,
must also have a very complex velocity field, according to the remark
by Meaburn, Lopez and Keir (1984) that a spherically expanding nebula
would not fit their dynamical data.
Our data do not show any velocity component clearly associated to the
post-merging expansion: the velocity of this latter might be close to the
dominant shell approaching/receding velocity, or very different but
undetected because of the much lower intensity of its emission (which
might have remained hidden within
the complex fit residuals shown in Fig.~\ref{atlas-resid}).
}

Another important result from the data presented is the existence of
significant gradients, typically with plane-parallel geometries rather
than spherical geometries. Also the mentioned ``shells" develop along
preferred directions, rather than along a radial direction: they should
better termed ``cylinders", although the physical mechanism leading to
this type of expansion geometry is not easy to understand. Nitrogen
ionization fraction also shows significant gradients: in the $\eta$ Car
shell these are explained by a polar-angle dependence of the ionizing
flux incident on the Nebula. In general, two preferred
directions appear to be those of the two obscuring dust lanes. In
addition to this dark material, also other layers of ``invisible"
material were shown to exist: one example is the massive (neutral?)
layer moving at RV$\sim -10$ km/s against which part of the Tr~14 shell
appears to have collided; another example is the
neutral material moving at RV$\sim -100$ to 0 km/s, causing the known
interstellar absorption lines towards Carina stars.
The complexity of observed structures in the studied region is such that
there are no truly representative ``average" conditions, holding to a
good approximation everywhere in the region: in order to put other
studies (like e.g.\ on the stellar populations of clusters Tr~14 and~16)
into a proper context, a careful consideration of local conditions is
highly needed.

\section{Summary}
\label{summary}

We have studied the profiles of optical emission lines of H$\alpha$, [N II], [S II]
and He I on more than 650 sightlines in the central part of the Carina Nebula.
We recover the known double-peaked line profiles in all lines, and
through gaussian fits to the lines we show that also single-peaked lines
actually arise from two distinct components, from approaching and receding
gas. The main properties of the gas yielding the bulk nebular emission are
the following:
\begin{itemize}
\item
{
The motion of gas in the central $20^{\prime} \times 20^{\prime}$ of the
Carina Nebula indicates the existence of
several distinct expanding shells, centered on
$\eta$ Car, Tr~14, WR25, and possibly the O3 star HD93250.}
\item The shape of these shells is non-spherical, with preferential
directions for expansion that are related to the geometry of the dark
obscuring dust lanes, and abrupt distortions likely caused by collisions
with higher-density material; traces of collisions between adjacent shells
are also found.
\item H$\alpha$ emitting gas is basically the same as He I emitting gas, while
significant kinematical differences are found with respect to gas emitting
in the forbidden lines.
\item The gas temperature is found to be $\sim 10^4$ K from line-width ratios
of H and He, and slightly hotter in the approaching component.
\item Gas densities were derived from the [S II] doublet ratio, and found
in the range 200-300 cm$^{-3}$, increasing towards the inner parts of
the Nebula, and near inter-shell boundaries.
\item The N ionization fraction is found to increase systematically
towards the South in the approaching component, with an almost opposite
trend for the receding component. In the $\eta$ Car shell,
this varying ionization is shown to be correlated with
angular distance from the Homunculus polar axis.
\item Within a small sample of sky directions, the differential Balmer
decrement between approaching and receding gas indicates a very low
extinction. Therefore, dust causing the occasionally large measured extinction
must be located either in front or behind the shells observed in
H$\alpha$ (and
stars contained within them) but not inside the shells.
\item The shell sizes and expansion velocities imply short timescales of 
$\sim 1.5 \cdot 10^5$ years for the ionized gas.
\end{itemize}

We also found significant residual emission at large absolute RVs, after
subtracting out the main gaussian components from the observed profiles.
{
A wide component is very often found in H$\alpha$, while narrow components are
occasionally found in [S II], and with less confidence in H$\alpha$ as well.
We have found several distinct pieces of evidence that such wide
H$\alpha$
wings arise from reflection of the $\eta$~Car spectrum (in agreement
with previous studies), and that the reflecting dust is likely radially
expanding to account for the observed asymmetry in these wings.
Reflection nebulosity is also indicated by the level of sky continuum
in places other than the Keyhole, notably near the Tr~14 core.
The [S II] narrow components are instead suggestive of localized shocks, as
also hypothesized for the diffuse X-ray emission already known in the region.
}

We remark that the observational data presented here were not originally
taken with the
purpose of studying the nebular emission: according to the Gaia-ESO
Survey observing strategy, a number of fibres is
routinely targeted at empty sky positions, in order to
eliminate the sky contribution from the stellar spectra.
This study demonstrates the rich scientific content of Gaia-ESO Survey
data, even surpassing their intended objectives.

\begin{acknowledgements}
We wish to thank an anonymous referee for his/her helpful suggestions.
Based on data products from observations made with ESO Telescopes at the
La Silla Paranal Observatory under programme ID 188.B-3002. These data
products have been processed by the Cambridge Astronomy Survey Unit
(CASU) at the Institute of Astronomy, University of Cambridge, and by
the FLAMES/UVES reduction team at INAF/Osservatorio Astrofisico di
Arcetri. These data have been obtained from the Gaia-ESO Survey Data
Archive, prepared and hosted by the Wide Field Astronomy Unit, Institute
for Astronomy, University of Edinburgh, which is funded by the UK
Science and Technology Facilities Council.
This work was partly supported by the European Union FP7 programme
through ERC grant number 320360 and by the Leverhulme Trust through
grant RPG-2012-541. We acknowledge the support from INAF and Ministero
dell' Istruzione, dell' Universit\`a' e della Ricerca (MIUR) in the form
of the grant "Premiale VLT 2012". The results presented here benefit
from discussions held during the Gaia-ESO workshops and conferences
supported by the ESF (European Science Foundation) through the GREAT
Research Network Programme.
This research has made use of the SIMBAD database,
operated at CDS, Strasbourg, France.
\end{acknowledgements}

\bibliographystyle{aa}

\onecolumn
\begin{table}[ht]
\centering
\caption{Double-gaussian fitting results for H$\alpha$. Full
table available electronically.} 
\label{fit-table-1}
\begin{tabular}{rllrrrrrrrr}
  \hline
Nr. & Id & Type/setup & RA & Dec & \multicolumn{6}{c}{H$\alpha$} \\
 & & & (J2000) & (J2000) & Norm$_{blue}$ & RV$_{blue}$ & $\sigma_{blue}$ &
                       Norm$_{red}$ & RV$_{red}$ & $\sigma_{red}$ \\
  \hline
1 & SKY\_\_10434580-5930497 & sky\_HR15N & 160.94083 & -59.51381 & 43435.14 & -23.61 & 15.61 & 58918.77 & 4.94 & 13.38 \\ 
  2 & SKY\_\_10445840-5933062 & sky\_HR15N & 161.24333 & -59.55172 & 18256.59 & -31.12 & 13.93 & 34321.84 & 5.23 & 15.21 \\ 
  3 & SKY\_\_10445040-5935467 & sky\_HR15N & 161.21000 & -59.59631 & 34377.49 & -28.47 & 13.77 & 65136.86 & -0.09 & 17.20 \\ 
  4 & SKY\_\_10443390-5934549 & sky\_HR15N & 161.14125 & -59.58192 & 34307.92 & -29.53 & 14.51 & 30563.08 & -15.04 & 23.81 \\ 
  5 & SKY\_\_10451730-5942205 & sky\_HR15N & 161.32208 & -59.70569 & 37527.72 & -28.25 & 15.50 & 22025.68 & 14.37 & 14.07 \\ 
  6 & SKY\_\_10451970-5945404 & sky\_HR15N & 161.33208 & -59.76122 & 24245.06 & -21.95 & 15.87 & 9017.24 & 11.42 & 15.95 \\ 
  7 & SKY\_\_10444430-5943333 & sky\_HR15N & 161.18458 & -59.72592 & 30519.80 & -34.15 & 14.82 & 53892.78 & 7.00 & 14.64 \\ 
  8 & SKY\_\_10440670-5947505 & sky\_HR15N & 161.02792 & -59.79736 & 8057.94 & -24.88 & 16.14 & 10603.72 & 10.00 & 15.99 \\ 
  9 & SKY\_\_10441789-5948152 & sky\_HR15N & 161.07458 & -59.80422 & 6677.84 & -21.83 & 17.05 & 5186.23 & 6.33 & 17.29 \\ 
  10 & SKY\_\_10441380-5943572 & sky\_HR15N & 161.05750 & -59.73256 & 20338.31 & -28.54 & 15.25 & 32785.90 & 2.83 & 16.33 \\ 
  11 & SKY\_\_10435880-5944037 & sky\_HR15N & 160.99500 & -59.73436 & 10317.80 & -29.29 & 13.93 & 37845.24 & 2.46 & 16.48 \\ 
  12 & SKY\_\_10434210-5933208 & sky\_HR15N & 160.92542 & -59.55578 & 57426.48 & -13.00 & 21.76 & 44337.46 & -0.38 & 14.19 \\ 
  13 & SKY\_\_10434580-5930497 & sky\_HR15N & 160.94083 & -59.51381 & 42161.66 & -24.58 & 15.38 & 65908.87 & 4.63 & 13.99 \\ 
  14 & SKY\_\_10452410-5932317 & sky\_HR15N & 161.35042 & -59.54214 & 9307.17 & -37.75 & 15.04 & 29908.25 & 3.90 & 17.32 \\ 
  15 & SKY\_\_10451230-5939301 & sky\_HR15N & 161.30125 & -59.65836 & 57393.14 & -32.90 & 14.24 & 46170.07 & 14.59 & 14.92 \\ 
  16 & SKY\_\_10454120-5937395 & sky\_HR15N & 161.42167 & -59.62764 & 25817.26 & -29.33 & 18.31 & 44071.29 & 6.19 & 14.04 \\ 
  17 & SKY\_\_10451730-5942205 & sky\_HR15N & 161.32208 & -59.70569 & 42171.85 & -28.09 & 15.39 & 24532.33 & 14.49 & 13.80 \\ 
  18 & SKY\_\_10451270-5940065 & sky\_HR15N & 161.30292 & -59.66847 & 43023.05 & -35.24 & 14.52 & 120581.13 & 6.87 & 14.57 \\ 
  19 & SKY\_\_10452060-5943215 & sky\_HR15N & 161.33583 & -59.72264 & 58355.08 & -25.32 & 16.09 & 24757.19 & 16.24 & 12.64 \\ 
  20 & SKY\_\_10451970-5945404 & sky\_HR15N & 161.33208 & -59.76122 & 24230.63 & -22.13 & 15.46 & 10209.32 & 10.43 & 16.47 \\ 
  21 & SKY\_\_10451060-5946127 & sky\_HR15N & 161.29417 & -59.77019 & 32485.39 & -22.23 & 15.18 & 10906.66 & 13.17 & 16.08 \\ 
  22 & SKY\_\_10444180-5944315 & sky\_HR15N & 161.17417 & -59.74208 & 36492.25 & -32.92 & 15.70 & 62097.00 & 8.54 & 14.32 \\ 
  23 & SKY\_\_10445440-5941005 & sky\_HR15N & 161.22667 & -59.68347 & 62865.35 & -34.23 & 14.74 & 38519.34 & 7.42 & 16.72 \\ 
  24 & SKY\_\_10435880-5944037 & sky\_HR15N & 160.99500 & -59.73436 & 9487.00 & -27.55 & 14.85 & 29200.92 & 3.55 & 16.34 \\ 
  25 & SKY\_\_10441810-5941175 & sky\_HR15N & 161.07542 & -59.68819 & 76112.44 & -29.23 & 14.03 & 135383.63 & 4.77 & 14.46 \\ 
   \hline
\end{tabular}
\end{table}
\begin{table}[ht]
\centering
\caption{Double-gaussian fitting results for [N II] 6584 and He~I 6678. Full table available electronically.} 
\label{fit-table-2}
\begin{tabular}{rrrrrrrrrrrrr}
  \hline
Nr. & \multicolumn{6}{c}{[N II] 6584} & \multicolumn{6}{c}{He I 6678} \\
 & Norm$_{blue}$ & RV$_{blue}$ & $\sigma_{blue}$ &
   Norm$_{red}$ & RV$_{red}$ & $\sigma_{red}$ &
   Norm$_{blue}$ & RV$_{blue}$ & $\sigma_{blue}$ &
   Norm$_{red}$ & RV$_{red}$ & $\sigma_{red}$ \\
  \hline
1 & 2943.68 & -31.66 & 14.71 & 11672.97 & 2.10 & 10.67 & 15235.06 & -22.16 & 13.79 & 18024.73 & 5.19 & 10.38 \\ 
  2 & 1130.41 & -24.70 & 15.34 & 4429.57 & 5.90 & 15.36 & 5555.79 & -30.83 & 11.55 & 11833.37 & 5.01 & 13.96 \\ 
  3 & 2528.38 & -22.84 & 13.49 & 3541.20 & 9.64 & 14.93 & 11447.99 & -28.24 & 10.67 & 22875.81 & 0.41 & 14.87 \\ 
  4 & 10198.94 & -25.17 & 12.26 & 3538.75 & -12.05 & 25.62 & 10756.85 & -29.39 & 12.29 & 11023.40 & -15.58 & 23.19 \\ 
  5 & 6328.36 & -28.63 & 13.16 & 2644.65 & 15.04 & 12.74 & 9441.61 & -27.95 & 11.69 & 7676.87 & 13.18 & 13.84 \\ 
  6 & 4011.71 & -17.86 & 13.23 & 2122.03 & -7.92 & 27.32 & 7065.39 & -17.61 & 13.23 & 1828.86 & 14.34 & 8.58 \\ 
  7 & 7207.63 & -33.96 & 12.22 & 5305.92 & 9.02 & 13.03 & 9951.92 & -32.18 & 14.02 & 17619.99 & 7.92 & 12.14 \\ 
  8 & 1587.50 & -20.77 & 13.62 & 937.43 & 11.35 & 15.09 & 1064.69 & -31.31 & 8.00 & 4338.92 & 5.87 & 17.68 \\ 
  9 & 1703.78 & -6.20 & 21.61 & 525.83 & -19.22 & 10.75 & 2466.90 & -10.28 & 18.70 & 266.07 & 16.04 & 6.46 \\ 
  10 & 1684.90 & -13.38 & 11.60 & 5422.81 & -7.82 & 21.49 & 5930.21 & -30.20 & 11.87 & 10660.00 & 2.94 & 13.60 \\ 
  11 & 4373.97 & -3.49 & 19.85 & 830.40 & -21.12 & 10.20 & 2127.89 & -33.45 & 8.29 & 12921.66 & 1.03 & 13.87 \\ 
  12 & 4685.66 & -33.46 & 14.92 & 6965.35 & 0.05 & 12.48 & 22855.47 & -8.70 & 18.91 & 10122.84 & 0.02 & 10.93 \\ 
  13 & 2800.86 & -33.03 & 14.29 & 13198.67 & 2.00 & 11.49 & 12023.08 & -26.25 & 12.42 & 23322.05 & 4.31 & 12.59 \\ 
  14 & 2186.93 & -16.49 & 17.16 & 3758.53 & 14.88 & 11.74 & 3443.08 & -32.95 & 15.86 & 7619.56 & 4.88 & 14.10 \\ 
  15 & 3167.91 & -31.88 & 13.19 & 5243.57 & 12.54 & 13.00 & 19613.70 & -32.19 & 11.88 & 15360.44 & 16.24 & 12.72 \\ 
  16 & 3011.29 & -30.13 & 18.32 & 8014.10 & 6.37 & 11.96 & 5516.82 & -33.65 & 14.53 & 16154.50 & 4.63 & 13.21 \\ 
  17 & 7134.41 & -28.55 & 13.13 & 3000.12 & 15.01 & 12.16 & 13148.70 & -28.02 & 12.73 & 8012.74 & 14.22 & 12.18 \\ 
  18 & 2994.32 & -29.87 & 14.65 & 12788.21 & 8.36 & 12.14 & 13590.73 & -35.56 & 11.32 & 43233.80 & 6.74 & 12.91 \\ 
  19 & 12502.16 & -26.42 & 12.22 & 3405.79 & 15.45 & 10.99 & 16662.09 & -24.51 & 12.98 & 8898.32 & 16.20 & 11.20 \\ 
  20 & 3188.13 & -20.08 & 11.56 & 3135.33 & -5.20 & 22.15 & 6146.43 & -22.07 & 12.88 & 3727.43 & 11.31 & 15.12 \\ 
  21 & 8912.84 & -21.74 & 12.62 & 1470.43 & 16.12 & 13.00 & 7698.10 & -22.79 & 12.53 & 3627.53 & 11.56 & 15.87 \\ 
  22 & 6138.80 & -29.72 & 12.10 & 4917.66 & 12.05 & 12.53 & 9370.82 & -33.21 & 13.37 & 22366.58 & 9.27 & 12.79 \\ 
  23 & 3765.67 & -24.34 & 19.62 & 4338.87 & 9.45 & 13.73 & 21464.20 & -33.51 & 12.20 & 12253.55 & 8.20 & 15.55 \\ 
  24 & 526.71 & -21.23 & 9.50 & 3724.93 & -3.80 & 20.34 & 3277.37 & -24.19 & 11.73 & 9583.29 & 4.69 & 13.13 \\ 
  25 & 11314.24 & -21.00 & 17.92 & 9176.47 & 8.85 & 12.04 & 25493.19 & -29.10 & 11.79 & 46905.75 & 4.43 & 12.16 \\ 
   \hline
\end{tabular}
\end{table}
\newpage
\begin{table}[ht]
\centering
\caption{Double-gaussian fitting results for [S II] 6717 and 6731. Full
table available electronically. Parameters are not given when the fit
did not converge.} 
\label{fit-table-3}
\begin{tabular}{rrrrrrrrrrrrr}
  \hline
Nr. & \multicolumn{6}{c}{[S II] 6717} & \multicolumn{6}{c}{[S II] 6731}
\\
 & Norm$_{blue}$ & RV$_{blue}$ & $\sigma_{blue}$ &
   Norm$_{red}$ & RV$_{red}$ & $\sigma_{red}$ &
   Norm$_{blue}$ & RV$_{blue}$ & $\sigma_{blue}$ &
   Norm$_{red}$ & RV$_{red}$ & $\sigma_{red}$ \\
  \hline
1 & 998.76 & -32.45 & 12.98 & 3398.53 & 4.20 & 10.39 & 727.14 & -33.94 & 12.95 & 2696.04 & 3.09 & 10.02 \\ 
  2 & 1655.72 & -4.40 & 19.00 & 736.73 & 9.41 & 9.97 & 1304.64 & -4.45 & 19.94 & 519.97 & 8.08 & 9.21 \\ 
  3 & 1733.36 & -20.27 & 18.14 & 950.03 & 14.14 & 12.06 & 1352.34 & -19.90 & 18.25 & 642.85 & 13.89 & 11.27 \\ 
  4 & 2687.79 & -21.74 & 12.12 & 740.18 & 7.11 & 32.20 & 2636.84 & -22.17 & 11.68 & 582.76 & 6.44 & 32.56 \\ 
  5 & 1748.67 & -25.49 & 12.72 & 836.83 & 16.73 & 11.97 & 1313.25 & -26.18 & 12.99 & 610.61 & 16.64 & 11.96 \\ 
  6 &  &  &  &  &  &  &  &  &  &  &  &  \\ 
  7 & 2213.11 & -31.97 & 11.54 & 1278.25 & 10.42 & 13.44 & 1662.81 & -32.62 & 11.37 & 933.80 & 9.52 & 13.01 \\ 
  8 &  &  &  &  &  &  &  &  &  &  &  &  \\ 
  9 &  &  &  &  &  &  &  &  &  &  &  &  \\ 
  10 &  &  &  &  &  &  &  &  &  &  &  &  \\ 
  11 & 469.30 & -18.53 & 9.54 & 1110.19 & -0.66 & 20.34 & 366.52 & -17.40 & 10.31 & 866.86 & -2.12 & 21.39 \\ 
  12 &  &  &  &  &  &  &  &  &  &  &  &  \\ 
  13 & 967.95 & -32.94 & 12.99 & 3802.68 & 4.36 & 11.14 & 691.68 & -33.68 & 12.85 & 3033.48 & 3.42 & 10.71 \\ 
  14 & 667.02 & -14.19 & 14.92 & 1498.09 & 16.50 & 11.59 & 448.63 & -17.26 & 13.95 & 1137.60 & 15.03 & 11.83 \\ 
  15 & 1093.46 & -28.97 & 11.44 & 1380.37 & 12.65 & 13.00 & 791.13 & -29.16 & 11.26 & 1049.22 & 11.93 & 12.59 \\ 
  16 & 1528.20 & -26.27 & 20.60 & 1249.28 & 10.53 & 10.68 & 1076.97 & -26.00 & 20.08 & 991.71 & 9.75 & 10.83 \\ 
  17 & 1988.85 & -25.27 & 13.06 & 960.64 & 17.27 & 11.89 & 1437.02 & -25.62 & 12.76 & 687.35 & 16.37 & 11.25 \\ 
  18 & 1013.07 & -27.23 & 13.68 & 2715.95 & 10.73 & 11.64 & 724.03 & -27.75 & 13.29 & 2280.19 & 9.77 & 11.25 \\ 
  19 & 4184.69 & -24.85 & 11.55 & 926.70 & 17.31 & 10.44 & 3029.70 & -25.83 & 11.14 & 699.69 & 15.96 & 10.98 \\ 
  20 &  &  &  &  &  &  &  &  &  &  &  &  \\ 
  21 &  &  &  &  &  &  &  &  &  &  &  &  \\ 
  22 & 3028.96 & -27.92 & 10.67 & 1108.90 & 14.41 & 12.80 & 2149.86 & -28.79 & 10.36 & 851.78 & 13.11 & 12.83 \\ 
  23 &  &  &  &  &  &  &  &  &  &  &  &  \\ 
  24 &  &  &  &  &  &  &  &  &  &  &  &  \\ 
  25 & 2519.56 & -30.03 & 15.18 & 2615.40 & 9.17 & 12.48 & 1890.12 & -30.80 & 15.31 & 2041.98 & 8.08 & 11.98 \\ 
   \hline
\end{tabular}
\end{table}

\begin{appendix}
\section{On the Giraffe/HR15N instrumental line profile}

{
In section~\ref{resid} we have examined best-fit residuals with
amplitudes of 1-2\% with respect to peak emission. Our ability to draw
meaningful conclusions from such low-level features (in relative terms)
in our data requires some detailed understanding of the instrumental
response. We have therefore examined Th-Ar calibration lamp spectra,
nearly simultaneous to some of the data studied here. These spectra are
crowded of lines, and therefore only very few lines were isolated and
bright enough for a study of their extreme tails down to less than 1\%
of peak. One such example is shown in Figure~\ref{calibr-lamp}, where
the abscissa $\Delta \lambda$ is transformed into a velocity scale to
facilitate comparison with results in section~\ref{resid}.
In the same Figure we show one-gaussian (red) and two-gaussian (blue)
best-fit models to the lamp line profile, and their residuals.
The one-gaussian model is discrepant from the observed line profile at a
level of 3-4\% maximum. The two-gaussian model residuals are much
smaller, slightly exceeding 1\% discrepancy. The profile obtained by
convolving the lamp profile with a gaussian with $\sigma=10$~km/s (e.g.,
a typical turbulent or thermal width), however, yields residuals of only
1\% when fitted with only one gaussian.
At face value, this would suggest that by approximating the instrumental
spectral response with a simple gaussian we should expect spurious fit
residuals of order of 1\% of peak emission.
However, in Figure~\ref{atlas-resid} we may observe that
the oscillating residuals around zero velocity often exceed the 1\%
level (indicated by the horizontal dashed lines), and show a typical
peak-to-peak RV interval of $\sim 30-40$~km/s, while the residual
oscillations in Fig.~\ref{calibr-lamp} occur at the 1-2 pixel level (RV
intervals $\sim 4-8$~km/s). The latter property especially would indicate
that the nebular fit residuals are unrelated to non-gaussianity of the
instrumental line profile.

To clarify the problem more accurately, we have considered the above
two-gaussian model for the lamp line profile as the basis for our
nebular-line fitting procedure.
The normalized lamp line profile was modeled using two gaussians as:

\begin{equation}
\lambda (v) =
\frac{A_1}{\sqrt{2 \pi \sigma_1^2}} \exp\left(-
\frac{(v-v_1)^2}{2\sigma_1^2}\right) +
\frac{A_2}{\sqrt{2 \pi \sigma_2^2}} \exp\left(-
\frac{(v-v_2)^2}{2\sigma_2^2}\right)
\end{equation}

with best-fit parameter values:
$v_1 = -6.404371$ km/s,
$\sigma_1 = 6.44256$ km/s,
$A_1 = 0.3966431$,
$v_2 = 3.454723$ km/s,
$\sigma_2 = 6.069641$ km/s, and
$A_2 = 0.6033569$.

The convolution of $\lambda(v)$ with a gaussian was then used in place
of a simple gaussian, as the basic ingredient of our two-component
nebular-line fits, as in
section~\ref{main}. The fit residuals were then computed as in
section~\ref{resid}. In this way we obtained residual patterns almost
indistinguishable from those shown in Fig.~\ref{atlas-resid}.
We conclude that all systematic peculiarities found in those residuals,
and discussed in section~\ref{resid}, do not arise because of
instrumental effects, but reflect actual properties of the nebular lines.

\begin{figure}
\resizebox{\hsize}{!}{
\includegraphics[bb=5 10 485 475]{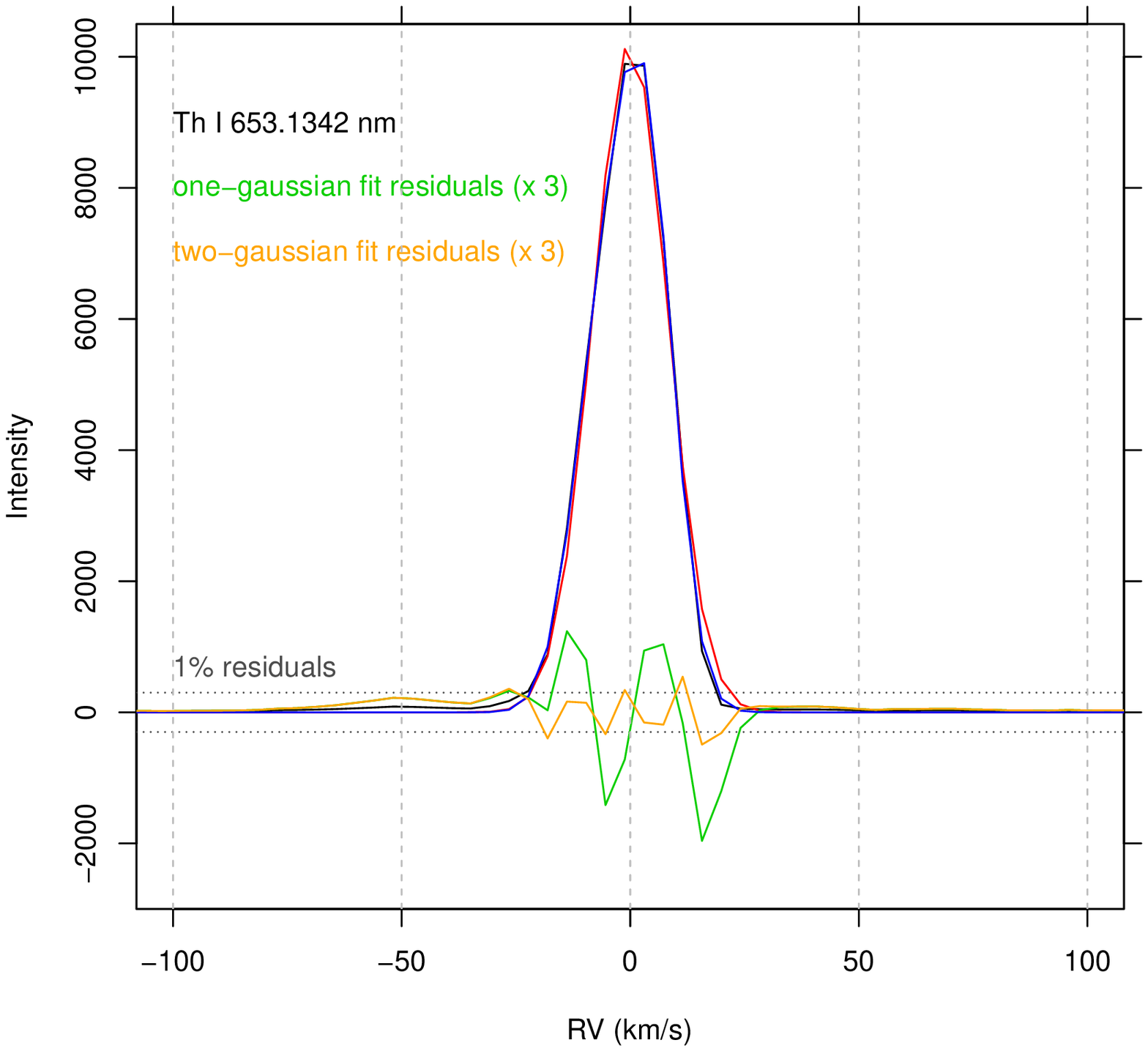}}
\caption{Giraffe HR15N Th-Ar lamp line (black), and best-fit
one-gaussian (red) and two-gaussian (blue) models. To facilitate
comparison with earlier results the abscissa is shown on a velocity
scale. Best-fit residuals, scaled-up three times, are shown in green and
orange for the one- and two-gaussian models, respectively.
The vertical dashed grey lines correspond to those shown in
Figure~\ref{atlas-resid}; the horizontal dotted lines indicate
amplitudes in the residuals of $\pm 1$\% of peak, like the horizontal
lines in Fig.~\ref{atlas-resid}.
\label{calibr-lamp}}
\end{figure}
}

\end{appendix}

\end{document}